\documentclass[iop]{emulateapj}

\usepackage{epsfig}
\usepackage{amsmath}
\usepackage{amssymb}
\usepackage{bm}

\usepackage{natbib}
\usepackage{color}
\usepackage{enumitem}
\usepackage{txfonts}

\renewcommand\appendixname{Appendices}

\begin{document}

\title{3D radiation non-ideal magnetohydrodynamical simulations of the inner rim in protoplanetary disks.}

   \author{M. Flock\altaffilmark{1,2}, S. Fromang\altaffilmark{2}, N. J. Turner\altaffilmark{1}, M. Benisty\altaffilmark{3}}
   \affil{$^1$Jet Propulsion Laboratory, California Institute of Technology, Pasadena, CA 91109, USA}
    \affil{$^2$Laboratoire AIM, CEA/DSM-CNRS-Universit\'e Paris 7,
  Irfu/Service d'Astrophysique, CEA-Saclay, F-91191 Gif-sur-Yvette,
  France}
    \affil{$^3$Universit\'e Grenoble Alpes, CNRS, IPAG, F-38000 Grenoble, France}
     \email{mflock@caltech.edu}

   \date{}

  \begin{abstract}
Many planets orbit within an AU of their stars, raising questions
about their origins.  Particularly puzzling are the planets found near
the silicate sublimation front.  We investigate conditions near the
front in the protostellar disk around a young intermediate-mass star,
using the first global 3-D radiation non-ideal MHD simulations in this
context.  We treat the starlight heating; the silicate grains'
sublimation and deposition at the local, time-varying temperature and
density; temperature-dependent Ohmic dissipation; and various initial
magnetic fields.

The results show magnetorotational turbulence around the sublimation
front at 0.5 AU.  The disk interior to 0.8 AU is turbulent, with
velocities exceeding 10\% of the sound speed.  Beyond 0.8 AU is the
dead zone, cooler than 1000 K and with turbulence orders of magnitude
weaker.  A local pressure maximum just inside the dead zone
concentrates solid particles, favoring their growth.  Over
many orbits, a vortex develops at the dead zone's inner edge,
increasing the disk's thickness locally by around 10\%.

We synthetically observe the results using Monte Carlo transfer
calculations, finding the sublimation front is near-infrared bright.  The models with net vertical magnetic fields develop
extended, magnetically-supported atmospheres that reprocess extra
starlight, raising the near-infrared flux 20\%.  The vortex throws a
non-axisymmetric shadow on the outer disk.  At wavelengths $>2\, \mu$m,
the flux varies several percent on monthly timescales.  The variations
are more regular when the vortex is present.  The vortex is directly
visible as an arc at ultraviolet through near-infrared wavelengths, given sub-AU spatial resolution.
\end{abstract}


\shorttitle{3D radiation non-ideal magnetohydrodynamical simulations of the inner rim.}
\shortauthors{Flock et al.}

   \maketitle

\section{Introduction and motivation}

Our grasp of planetary systems' origins relies on our understanding of
the disks of gas and dust found orbiting young stars. Here we focus on
the environment for planet formation in the disks' hot central region
where temperatures exceed 1000~K. Among the main processes governing the
temperatures in this region are the sublimation and deposition of
silicate grains, which change the opacity by orders of magnitude
\citep{pol94}. %% 
Since the starlight is a major source of heating, the opacity has a
major impact on temperatures. 

A key process governing the turbulent
stirring of the planet-forming materials is magneto-rotational
instability \citep{bal91}. Magnetic fields are coupled to the plasma,
and the instability converts gravitational potential energy into the
kinetic and magnetic energy of turbulence
\citep{jin96,gam96,mil00,hir11}, when temperatures are high enough for
thermal ionization \citep{ume88,des15}. The decline in temperature
with distance from the star thus leads to a decline in the magnetic
stresses across the thermal ionization threshold. This means the
surface density increases across the threshold if the inflow is in
steady-state, so that the mass flow rate is independent of
distance. The resulting local pressure maximum can concentrate solid
particles in the size range where their stopping time due to gas drag
is comparable to the orbital period
\citep{hag03,lyr08,kre09,dzy10,lyr12,fau14b}, possibly allowing for in
situ planet formation \citep{cha14}. 

In addition, the pressure maximum can act as a planet trap: young
planets' inward migration under their tidal interaction with the disk
comes to a halt near the pressure peak
\citep{mas06,mat09,kre12,bit14,hu15}. Concentrating both pebbles and
protoplanets in a region where dynamical timescales are short has the
potential to lead quickly to the growth of planets. 

Of all young stars, perhaps the best suited for measuring the disks' hot central
regions are the Herbig stars, which have masses a few times the
Sun's. Nearby examples are bright enough to be observed with
near-infrared interferometry down to the angular scale of the disk's
inner rim \citep{dul10,kra15} yielding maps of the silicate
sublimation front \citep{ben11}. However these objects' spectra show puzzlingly large
near-infrared fluxes \citep{hil92,chi01,mil01,mee01,vin06}. Radiation
hydrostatic \citep{mul12} and radiation hydrodynamic models
\citep{flo16} produce too little flux at wavelengths 2-4~$\mu$m by
factors up to several. Ingredients modelers must consider include the
transfer of the starlight into the disk, and the escape of the
re-radiated infrared emission; the sublimation and deposition of the
dust grains that provide most of the opacity; and the forces
supporting the disk material against the star's gravity
\citep{kam09}. Models can yield near-infrared excesses closer to the
observed range if some disk material near the sublimation front is
either launched into a wind \citep{ban12} or supported on the magnetic
fields escaping from MRI turbulence within the disk \citep{tur14a}. In
both pictures, the magnetic forces lift some material above its
hydrostatic position, increasing the height where the starlight is
absorbed, so that a bigger fraction of the stellar luminosity is
reprocessed into thermal emission at distances where the emission
comes out at near-infrared wavelengths. Until now, there was no global
modeling of magneto-rotational turbulence at the inner rim, treating
the dust sublimation and radiation transfer together with the MHD. 

In this work we present the first 3-D radiation non-ideal MHD
simulations of protostellar disks to include starlight heating,
silicate grains' sublimation and deposition at the local temperature
and density, and Ohmic dissipation depending on the thermal
ionization. We test the results against various observational
constraints, comparing the spectral energy distributions, images, and
lightcurves at different wavelengths. The models let us address
several important questions: How does the MHD turbulence affect the
sublimation front? What are the consequences for the
starlight-absorbing surface, the system's near-infrared emission, and
its time variability? And what controls the dynamics at the location
where the solids are concentrated? 

We describe in section~2 the radiation MHD methods and the treatment
of the dust sublimation and deposition. Section~3 deals with the
results of the calculations, and section~4 with the comparison against
observations. We discuss the implications in section~5, and summarize
our conclusions in section~6. 

\section{Methods and setup}
\label{sec:method}

In this section we briefly summarize the method and the setup of the
3D radiation non-ideal MHD simulations. The relevant equations are given in Section~\ref{sec_eq}. The
resistivity module which determines the magnetic field coupling
parameter is presented in Section~\ref{sec_res}. The initial
and boundary conditions are presented in Sections~\ref{sec_in} and~\ref{sec_bound}. For full details off the radiation transfer
and the dust evaporation modules, we refer the reader to our previous
works \citep{flo13,flo16}.

\subsection{Radiation non-ideal MHD equations}
\label{sec_eq}
In this paper, we solve the following radiation non-ideal MHD equations in a
spherical coordinate system $(r,\theta,\phi)$:
\begin{eqnarray}
 \frac{\partial \rho}{\partial t } + \nabla \cdot \left [  \rho \bm{v}\right ] &=&
0 \, , \label{eq:MDH_RHO} \\
\frac{\partial \rho \bm{v}}{\partial {t}} + \nabla \cdot \left [
  {\rho} \bm{v} \bm{v}^T - \bm{B}\bm{B}^T \right ] + \nabla {P_t}
&=& - {\rho} \nabla {\Phi} \, , \label{eq:MDH_MOM} \\
\frac{\partial {E}}{\partial {t}} + \nabla \cdot \left [ ( {E} +
  {P_t})\bm{v} - (\bm{v}\cdot\bm{B})\bm{B} \right ]  &=& - {\rho}
\bm{v} \cdot \nabla {\Phi} - \nabla \cdot {F}_* \nonumber \\& &  - \kappa_{P} {\rho} {c} ( {a_R} {T}^4 - {E_R} )\nonumber\\ & & 
-\nabla \cdot \left [ ({\eta} \cdot \bm{J}) \times \bm{B} \right ], \label{eq:MDH_EN} \\
\frac{\partial {E_R}}{\partial {t}} - \nabla \frac{ {c} {\lambda}}{ \kappa_{R} \rho} \nabla {E_R} &=& \kappa_{P} {\rho} {c} ( {a_R} {T}^4 - {E_R}) \, , \label{eq:ER} \\
\frac{\partial \bm{B}}{\partial {t}} - \nabla \times (\bm{v} \times
\bm{B}) &=& -\nabla \times({\eta} \cdot \bm{J}) \, , \label{eq:MDH_MAG} 
\end{eqnarray}
where ${\rho}$ is the density, $\bm{v}$ is the velocity and $\bm{B}$
is the magnetic field\footnote{The magnetic field already includes the
normalization factor $1/\sqrt{4\pi}$.}. $\bm{J}= \nabla \times \bm{B}$
is the current density and ${\eta}$ is the tensor magnetic resistivity. 
The total pressure is given by ${P_t} = {P} + \bm{B}^2/2$
where the gas pressure relates to the temperature $T$ through 
\begin{equation}
{P}=  \frac{{\rho} {k_B} {T}} { {\mu_g} {u}} \, ,
\end{equation}
where ${\mu_g}$ is the mean molecular weight, ${k_B}$ is 
the Boltzmann constant and ${u}$ is the atomic mass unit. $T$ stands for the temperatures of the gas and dust which are assumed equal thanks to collisional exchange of thermal energy at these high gas densities.

Introducing
the gravitational constant $G$, the gravitational potential is
calculated according to 
\begin{equation}
{\Phi} = {G} {M_*}/ {r} \, ,
\end{equation}
where ${M_*}$ is the stellar mass. ${E}$ denotes the total energy and
is given by the relation ${E}= {\rho} {\epsilon} + 0.5 {\rho} \bm{v}^2
+ 0.5 \bm{B}^2$ where ${\rho} {\epsilon}=P/(\Gamma-1)$ is the
gas internal energy, $\Gamma$ being the adiabatic index. The radiation
energy is denoted $E_{R}$ while ${F}_*$ stands for the
frequency-integrated irradiation flux. 
 $F_*$ is calculated as
\begin{equation}
F_*(r) = \left (  \frac{R_*}{r}\right )^2  \sigma_b T_*^4 e^{-\tau_*}, 
\label{eq:IRRAD}
\end{equation}
with the Stefan-Boltzmann constant $\sigma_b$, the stellar surface temperature $T_*$ and the radius $R_*$ of the star. The radial optical depth of the irradiation flux $\tau_*$ is calculated using the opacity at the stellar temperature $\kappa_P(T_*) = 2100\, \rm cm^2.g^{-1}$, see also Section 2.1 in \citet{flo16}. ${\kappa_R}$ and ${\kappa_P}$ are the Rosseland and Planck mean opacity,
respectively. Both opacities include gas and dust contributions. The dust opacity is set to $\kappa_P(T_{ev}) = 700\, \rm cm^2\, g^{-1}$ which represents the opacity at the dust sublimation temperature per gram of dust. The gas opacity is set constant to $10^{-4} \rm cm^2\, g^{-1}$ which represents the opacity per gram of gas. For more details on the opacities we refer to our previous work \citep{flo16}. Finally, ${ a_R}$ is the radiation constant and ${c}$ stands for the speed of light. 

The gas is a mixture of hydrogen and helium with solar abundance
\citep{dec78} so that $\mu_g =2.35$ and $\Gamma=1.42$. For the typical
density and temperature we considered, most of the hydrogen is 
bound in molecular hydrogen\footnote{We have checked that for $\rho <
10^{-14}$ g.cm$^{-3}$ and $T < 1500 K$ over 50 \% of the hydrogen is bound in
H$_2$.}. Silicates dust grains are present when the
temperature is smaller than a critical temperature noted $T_{ev}$ and
sublimates otherwise. As in our previous paper \citep{flo16}, we
follow \citet{pol94} and \citet{ise05} and determine $T_{ev}$ according
to 
\begin{equation}
T_{ev}=2000\, K \left ( \frac{\rho}{1 g\, cm^{-3}} \right )^{0.0195}.
\label{eq:ev}
\end{equation}
$T_{ev}$ is then used to calculate the local dust density $\rho_d$,
assuming perfect mixing of dust and gas \citep[see Eq.(10)
in][]{flo16}. 
%
%We use the same dust opacity as in our previous papers:
%the mean opacity of the dust for the absorption of stellar radiation
%amounts to $\kappa_P(T_*) = 2100\, \rm cm^2.g^{-1}$ while
%$\kappa_P(T_{ev}) = 700\, \rm cm^2\, g^{-1}$ represents the cooling
%efficiency at the dust sublimation temperature. 
%
%For further details on the
%irradiation flux, the dust evaporation module, the physical constants
%and the opacity, we refer the reader to our previous paper
%\citep[][section 2.1 and 2.2 therein]{flo16}.

\subsection{Resistivity}
\label{sec_res}

We implemented a simple treatment of the resistivity in order to model
the ionization transition at 1000~K \citep{ume88,des15}. To do so, we set
\begin{equation}
\eta= \frac{c_s^2}{\Omega Re_m} \, , 
\end{equation}
where the magnetic Reynolds number $Re_m$ dependence on the
temperature is given by
\begin{equation}
Re_m = 5 \times 10^4  \left ( 1-\tanh{\left ( \frac{1000-T}{25} \right
  )} \right ) \, . 
\label{eq:rem}
\end{equation}
The asymptotic value of the magnetic Reynolds number $Re_m = 10^5$ is
used for temperatures above 1000~K. 
This upper limit is high enough to obtain sustained MRI turbulence closely resembling the ideal-MHD limit \citep{flo12}.  
For temperatures below 1000~K the Reynolds number decreases and we set the lower limit of the magnetic Reynolds number to $Re_m^{DZ}=1$. Such a value is low enough to ensure the damping of the MRI inside the dead-zone (Elsasser number $\ll 1$). 

\subsection{Initial conditions}
\label{sec_in}

\begin{figure}
\centering
%  \hspace{-1.3cm}
%\begin{minipage}{0.4\textwidth}
\resizebox{\hsize}{!}{\includegraphics{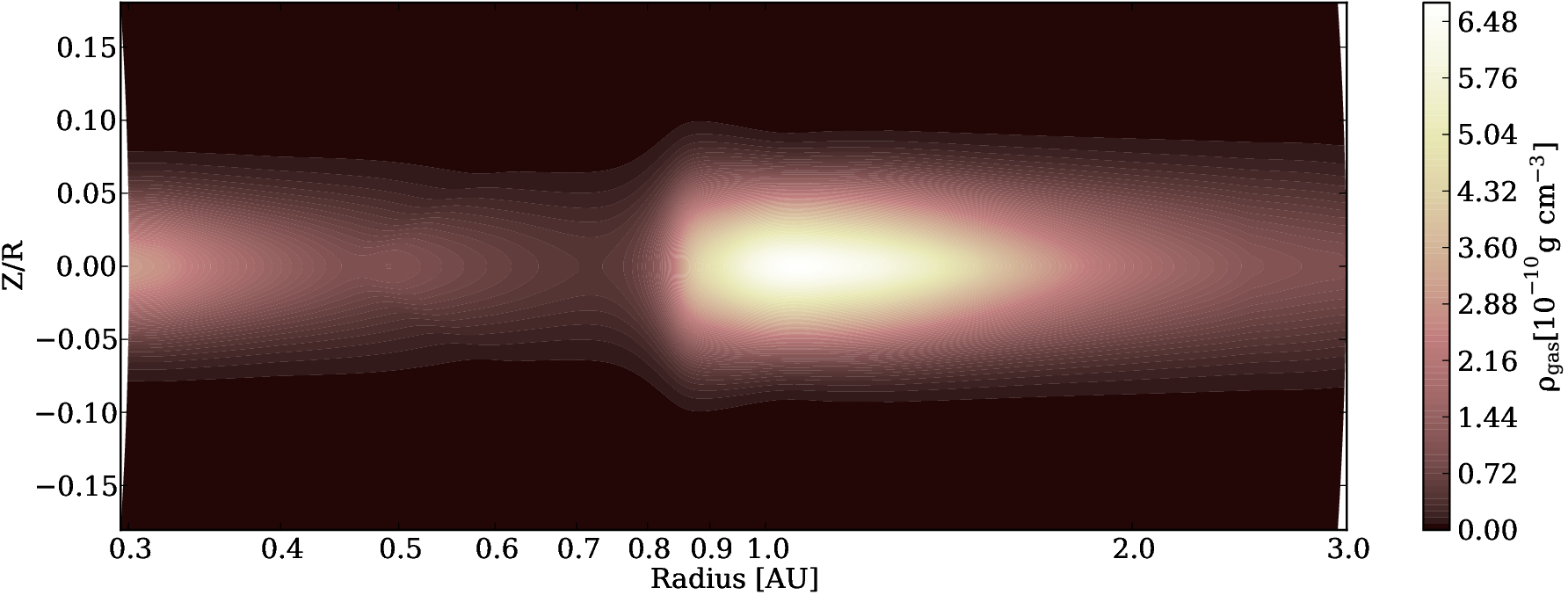}}
\resizebox{\hsize}{!}{\includegraphics{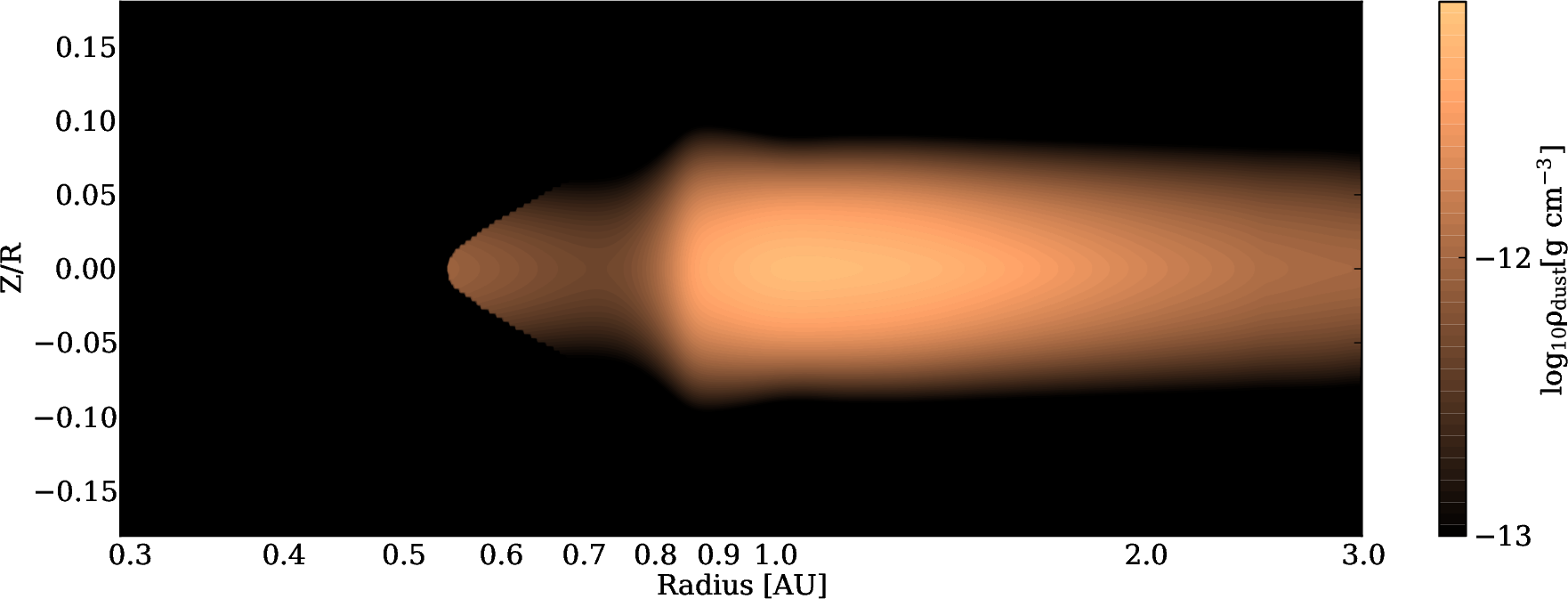}}
\resizebox{\hsize}{!}{\includegraphics{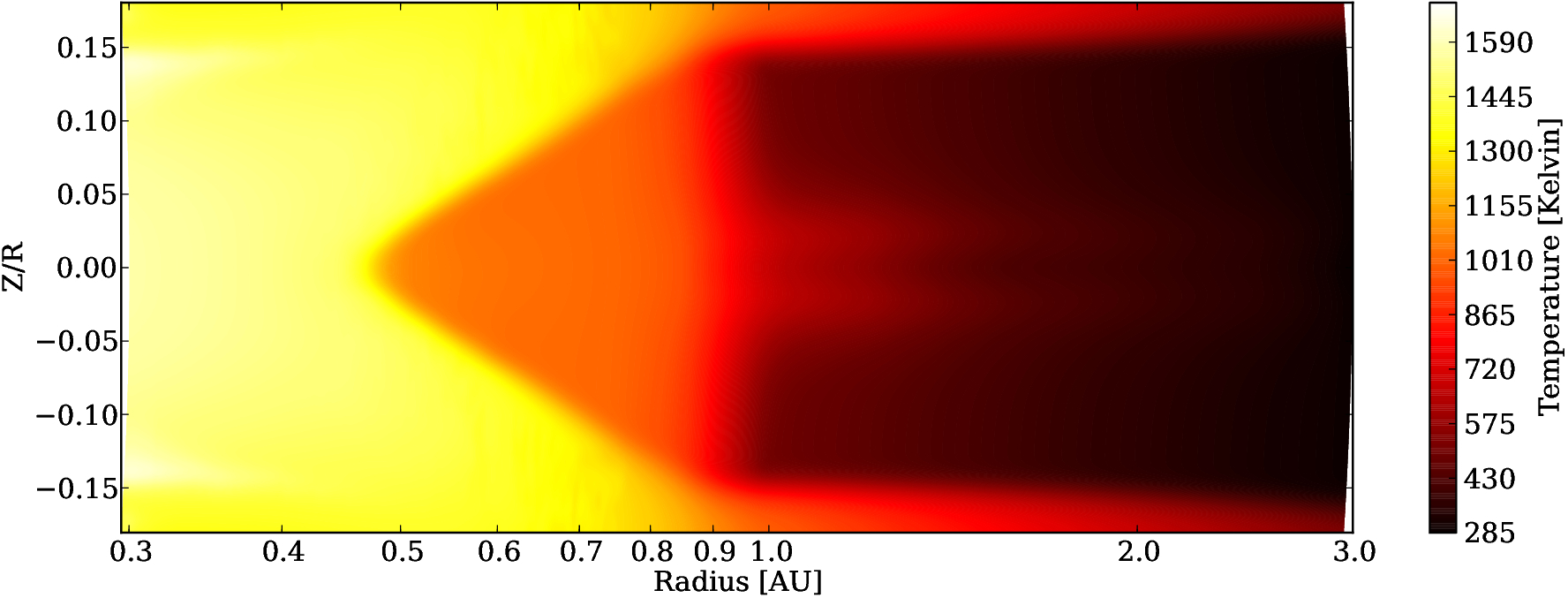}}
\resizebox{\hsize}{!}{\includegraphics{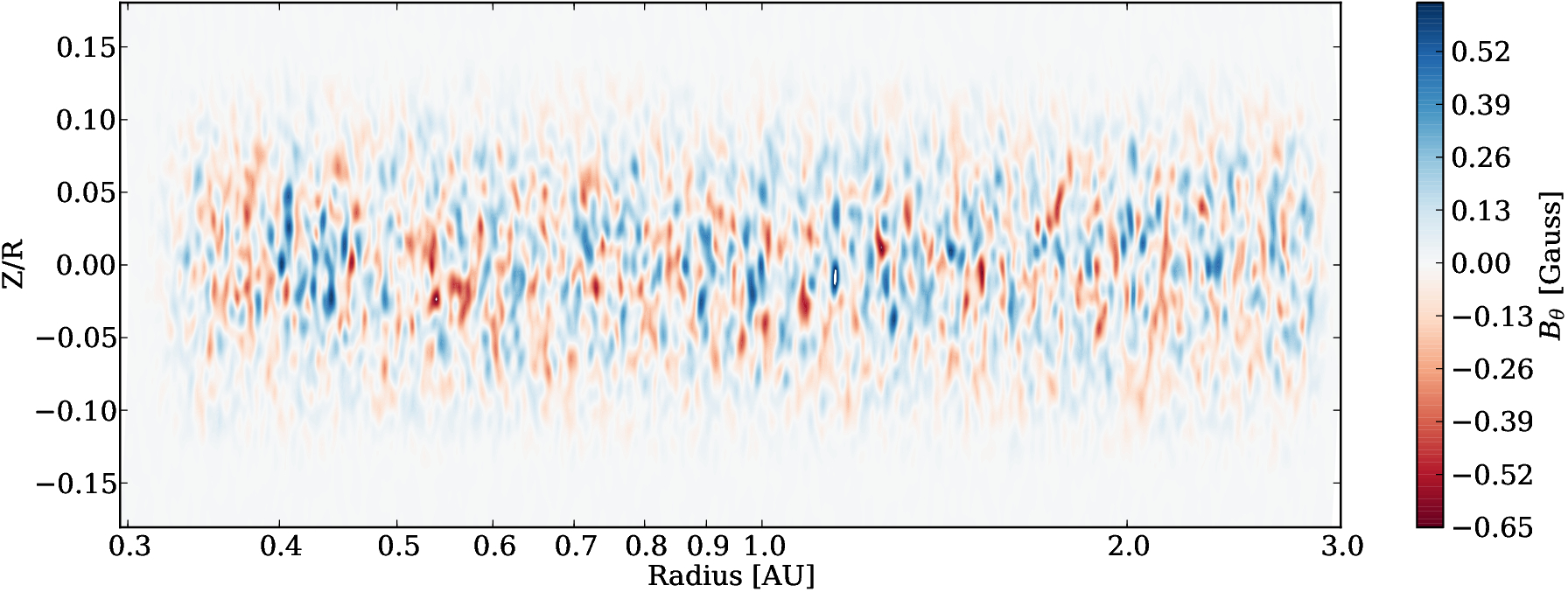}}
% \end{minipage}
\caption{Initial profile in the $R$-$Z/R$ plane for the gas density
  (top), the dust density (second row) and the temperature (third
  row). Those are snapshots taken from 2D radiation viscous
  hydrodynamical simulations in steady-state. Bottom: Meridional slice
  of the initial $B_\theta$ magnetic field component, generated from
  the vector potential $A_r$.} 
\label{fig:hdinit}
\end{figure}

In order to initialize the simulations presented here, we used the
final snapshot of the axisymmetric radiation viscous 2.5D\footnote{2.5D represents
the use of 2 dimensions ($r$,$\theta$) and 3 components
($r$,$\theta$.$\phi$) for the velocity and magnetic field vectors.}
hydrodynamical simulation \texttt{RMHD\_1e-8} of \citet{flo16}, for
which the flow has reached a steady-state characterized by a uniform
accretion rate of $\dot{M}=10^{-8}$ solar mass per year. We refer the
reader to \citet{flo16} for more details on that particular
simulation, and only show here the resulting 2D distribution of
$\rho$, $\rho_d$ and $T$ in the disk meridional plane (Fig.\ref{fig:hdinit}). These 2D fields are extended in the azimuthal
direction to cover the range $[0,0.4]$ radian and $[0,1.6]$ radian for
the models \texttt{RMHD\_P0\_4} and \texttt{RMHD\_P1\_6}, respectively.  
%% Here, we only remind that 
%% these simulations capture the process of accretion heating, which in
%% particular has an effect on the temperature spatial profile. 

To trigger the MRI, we investigate two different configurations for the
magnetic field geometry at the beginning of the simulations. First, a
random zero-net flux magnetic field is used for the models
\texttt{RMHD\_P0\_4} and \texttt{RMHD\_P1\_6}. A snapshot of the
initial magnetic field is shown in Fig. \ref{fig:hdinit} (bottom
panel). For model \texttt{RMHD\_P0\_4\_BZ}, we also added a vertical
net flux magnetic field. In Appendix~\ref{ap_1}, we detail the
procedure we used to generate the magnetic vector potential in both
cases.  The naming convention of the radiation MHD (RMHD) models includes the size of the azimuthal domain given in radians (e.g. \texttt{P0\_4} for $\Phi_{max}=0.4$) and if a vertical magnetic field is included, \texttt{BZ}. The model parameters are summarized in Table~\ref{tab:info}.
\begin{table}
\begin{tabular}{lll}
\hline
$\dot{M}$ & $10^{-8}$ $\rm M_\sun/year$ \\
Stellar parameter & $T_*=10000\, \mathrm{K}$, $R_*=2.5\, R_\sun$, $M_*=2.5\, M_\sun$\\
Opacity & $\kappa_P(T_*)=2100\, \rm cm^2\, g^{-1}$\\ 
        & $\kappa_P(T_{ev}) = 700\, \rm cm^2\, g^{-1}$\\ 
        & $\kappa_{gas} = 10^{-4}\, \rm cm^2\, g^{-1}$\\
Dust to gas mass ratio     & $f_0=0.01$\\
\hline
Cell aspect ratio & $R \Delta \theta/ \Delta R : R \Delta \Phi  / \Delta R \sim 1.1 : 1.2$ \\
$R_{in}-R_{out} : Z/R $ & 0.3-3AU : $\pm 0.36$\\ 
$N_r \times N_\theta$ & 896 x128\\ 
\texttt{RMHD\_P0\_4} & $\Phi_{max}=0.4$, $N_\phi=128$, runtime 300 orbits\\
\texttt{RMHD\_P1\_6} & $\Phi_{max}=1.6$, $N_\phi=512$, runtime 150 orbits\\
\texttt{RMHD\_P0\_4\_BZ} & $\Phi_{max}=0.4$, $N_\phi=128$, runtime 70 orbits\\
\hline
\end{tabular}
\caption{Model parameter for the radiation MHD model \texttt{RMHD\_P0\_4}, \texttt{RMHD\_P1\_6} and model \texttt{RMHD\_P0\_4\_BZ}.}
\label{tab:info}
\end{table}

\subsection{Boundary conditions and buffer zones}
\label{sec_bound} 

In the radial direction we use zero gradient conditions for all
variables while $v_r$ is set to enforce vanishing mass inflow. In the meridional direction we extrapolate the logarithmic
density and the temperature in the ghost cells. The ghost cells are a set of additional cells at the domain boundary which provide the boundary values for the integration. For the velocity
$v_\theta$ we set a zero mass inflow condition. The azimuthal boundaries are periodic. In addition, we use a buffer zone at the radial inner
boundary over the radial range $[0.3,0.35]$ AU to avoid effects
arising from the presence of the boundary. In this zone we damp the
radial and vertical velocities and increase linearly the magnetic resistivity
in order to reach a magnetic Reynolds number of $10$ at the location
of the inner radial boundary. In addition, we used the
same modified gravitational potential inside the buffer zone as used
in the 2D radiation hydrodynamical simulations \citep[see Appendix E
in][]{flo16}. This last modification affects the region with $|Z/R| > 
0.1 $ and $R < 0.35 AU$, and manifests itself as a layer with
temperatures slightly larger than expected (see white zone in
Fig.~\ref{fig:hdinit}, third panel). R and Z represent here the cylindrical coordinates. This buffer zone is excluded from
the analyses presented below. 

\subsection{Diagnostics}

We determine the strength of the turbulence by calculating the stress
to pressure ratio $\alpha$:
\begin{equation}
\rm \alpha = \frac{ \int \rho \Bigg( \frac{T_{r \phi}}{P} + \frac{M_{r
      \phi}}{P} \Bigg)dV} {\int \rho dV} = \frac{ \int \rho \Bigg(
  \frac{\rho v'_{\phi}v'_{r}}{P} - \frac{B_{\phi}B_{r}}{P}\Bigg)dV}
    {\int \rho dV}, 
\label{eq:ALPHA}
\end{equation}
which is the sum of the Reynolds stress $\rm T_{r \phi}$ and Maxwell
stress $\rm M_{r \phi}$\footnote{We compared the mass weighted
  integral for the $\alpha$ parameter with the classical definition
  $\alpha=\int (T_{r \phi} +  M_{r \phi}) dV/ \int P dV$. The radial
  profiles are almost identical with maximum relative deviations of
  around 5 \% close to the inner rim.}. Radial profiles of $\alpha$
are obtained by integration along $\theta$ and $\phi$. 2-D profiles are obtained by integration along $\phi$.  

We note that the time units are inner
orbits, which always refer to the orbital time at 0.3 AU. 

\section{Results}

\begin{figure}
  \centering
  \resizebox{\hsize}{!}{\includegraphics{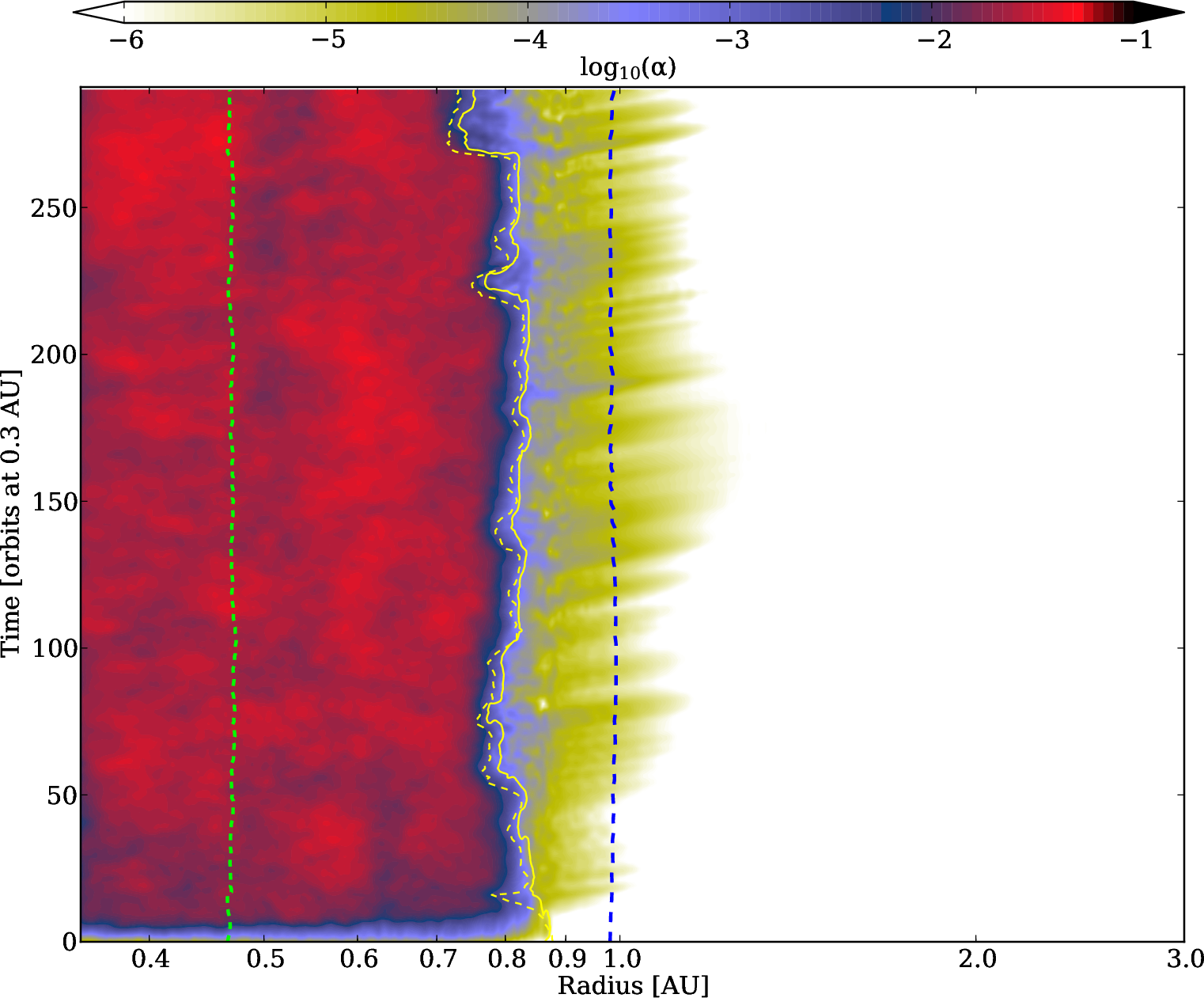}}
\caption{Time evolution of the stress to pressure ratio $\alpha$ over
  radius for model \texttt{RMHD\_P0\_4}. Midplane positions are shown
  for (from left to right) the inner rim (green dashed line), the 900~K temperature
  contour (yellow dashed line), the inner dead-zone edge
  with $\Lambda = 1.0$ (yellow solid line) and the dust concentration radius
  (blue dashed line).} 
\label{fig:alpha}
\end{figure}

\begin{figure}
  \centering
  \resizebox{\hsize}{!}{\includegraphics{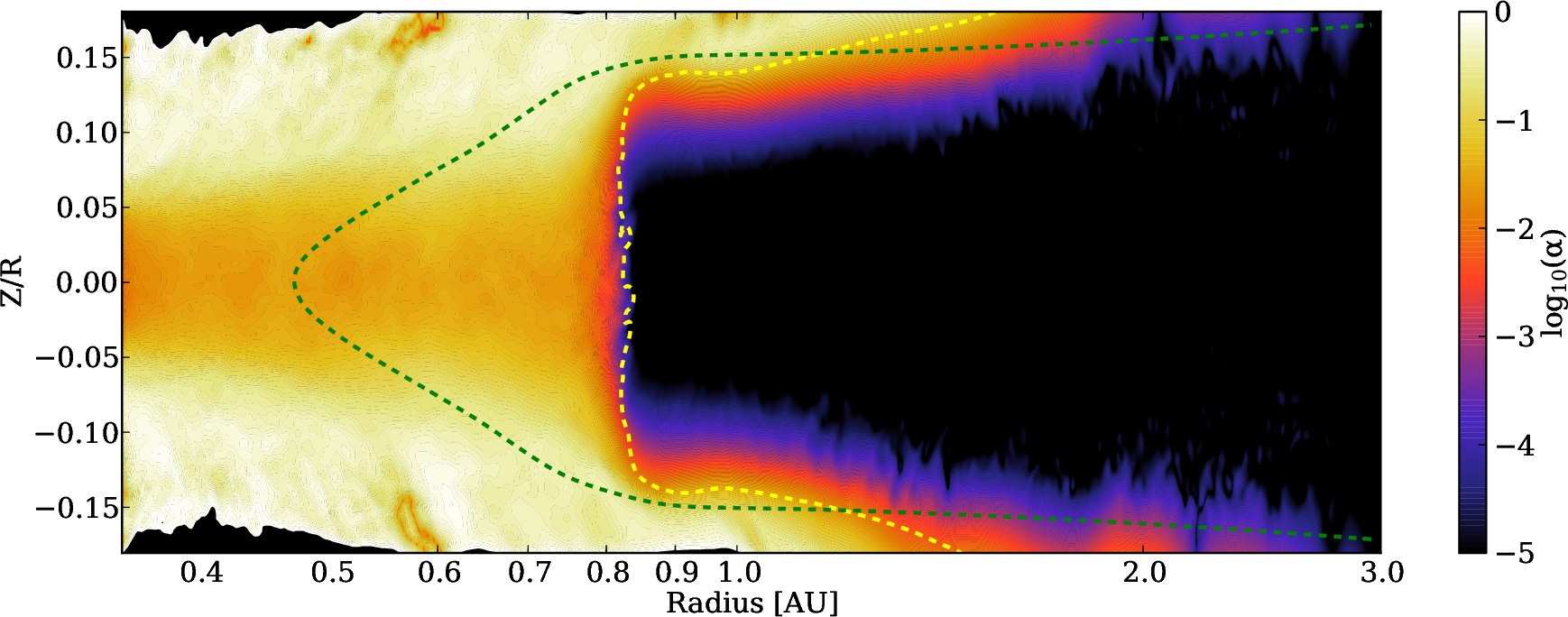}}
\caption{2D profile of the stress to pressure ratio $\alpha$ in the
  $R$-$Z/R$ plane for model \texttt{RMHD\_P0\_4}. Time average is done
  between 20 and 300 inner orbits. Overplotted are the contours for
  the inner rim (green dashed line) and the inner dead-zone edge
  (yellow line) ($\Lambda=1.0$).} 
\label{fig:al2D}
\end{figure}

We focus our analysis on the dynamics of three distinct regions of the
disk. The first is the inner rim, which we define as the
irradiation $\tau_* =1$ line. The second is the inner edge of the dead-zone, defined as the region where the Elsasser
number $\Lambda = v_a^2/(\eta \Omega)$ ($v_a$ being the
Alfv\'{e}n velocity) is smaller than unity. MRI turbulence is damped when $\Lambda$ is less than unity \citep{san02II}. The third is the dust concentration radius, and is defined as the location when pressure reaches a maximum and where the gas azimuthal
velocity is exactly Keplerian. The gas is rotating with
super(sub)-Keplerian velocities inward (outward) of that
location. We will focus on those three regions in the
following analysis and mark their positions in Figs.~\ref{fig:alpha} to~\ref{fig:turbq_vert}.   

We first investigate the detailed disk dynamics using the results of
model \texttt{RMHD\_P0\_4} in Sections~\ref{sec:dyn} and~\ref{sec:timeav}. The results of model
\texttt{RMHD\_P0\_4\_BZ} are presented in Section~\ref{sec:vert}. The
question of whether large-scale and long-term non-axisymmetric
perturbations arise in the simulations is investigated using model
\texttt{RMHD\_P1\_6} in Section~\ref{sec:large}.   

\subsection{Time evolution} 
\label{sec:dyn}

Fig.~\ref{fig:alpha} shows the time variations of $\rm \alpha(R)$ for
model \texttt{RMHD\_P0\_4}. The immediate result is that the model
quickly reaches a quasi-steady state. As a result of the MRI, the flow
is turbulent in the inner disk ($r<0.8$ AU) during the entire
simulation, with typical values of $\rm \alpha \sim 0.03$. Such a rate
for the flux of angular momentum is in agreement with previous locally isothermal
global ideal MHD simulations \citep{fro06,flo11,par13} and global
radiation ideal MHD simulations \citep{flo13}. 

The inner rim ($\tau_* =1$) in the midplane is located at around $0.47$ AU inside this highly turbulent
region (green dashed line). Its position remains almost unaffected by the turbulence because the optical depth to the star is the line integral through a large number of uncorrelated turbulent density fluctuations. The dead-zone inner edge is 
located at around $0.8$ AU (yellow solid line). At this position, the
Elsasser number $\Lambda$ drops below unity. In our model, the
dead-zone inner edge is also close to the temperature contour $T \sim
900$~K (yellow dashed line) and corresponds to the location where the magnetic Reynolds 
number drops below $\sim 100$. At this position, the value of $\alpha$
drops by several orders of magnitude. We note that the exact radial
position of the dead-zone inner edge shows variations of around 2 to 3 disk
scale heights in the radial direction with time (with typical displacement of
the order of $\Delta R \sim$ 0.1 AU). This is due to temperature
variations associated with the turbulence. Finally, the dust concentration radius is
located inside the dead-zone at around $1$ AU (blue dashed line) and
displays little sign of radial variations. 

\begin{figure}
  \centering
  \resizebox{\hsize}{!}{\includegraphics{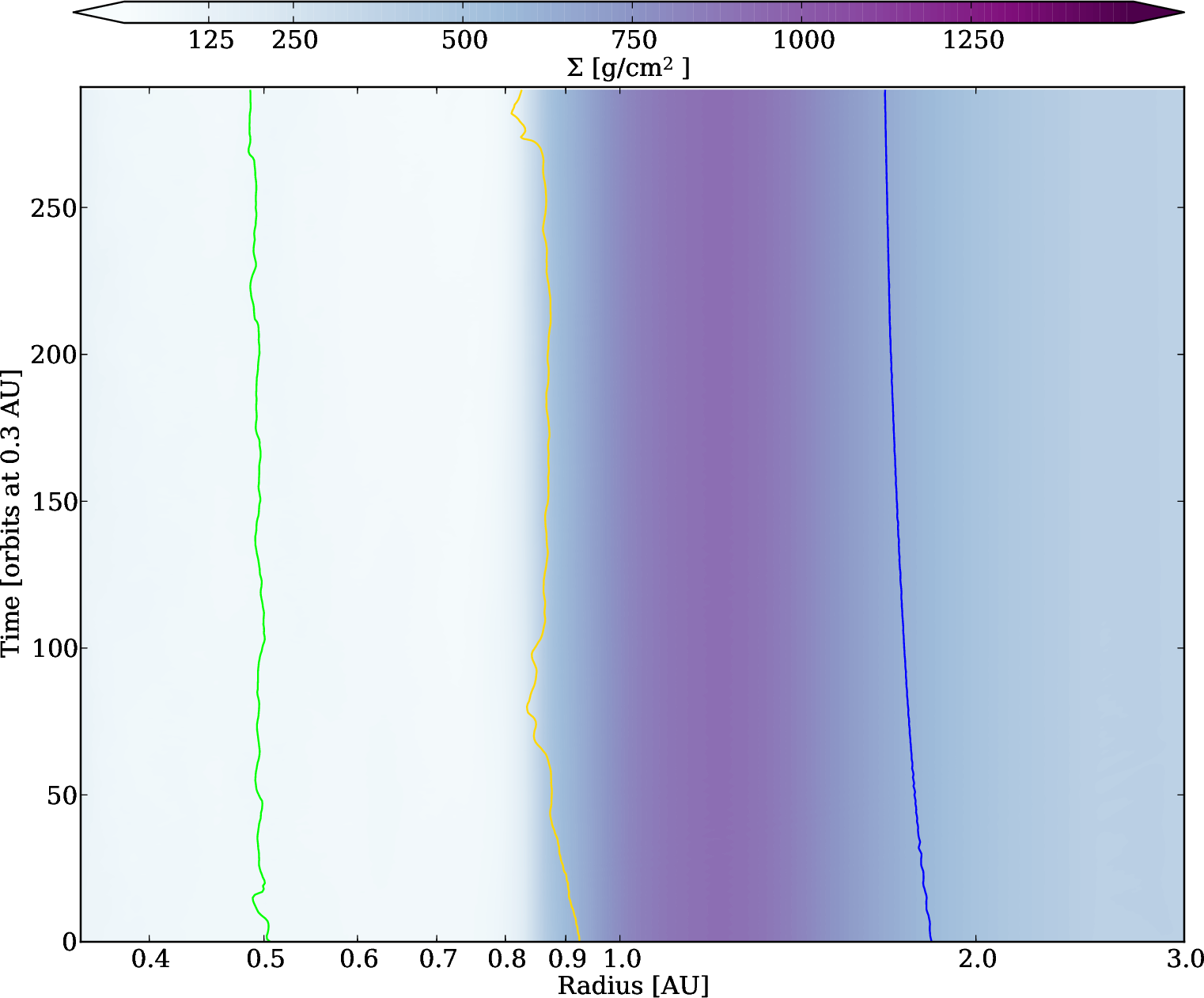}}
\caption{Time evolution of the surface density over
  radius for model \texttt{RMHD\_P0\_4}. Temperature contours are shown
  for 1200~K (green line), 800~K (yellow line), and 400~K (blue line).} 
\label{fig:surftime}
\end{figure}

\begin{figure}
  \centering
  \resizebox{\hsize}{!}{\includegraphics{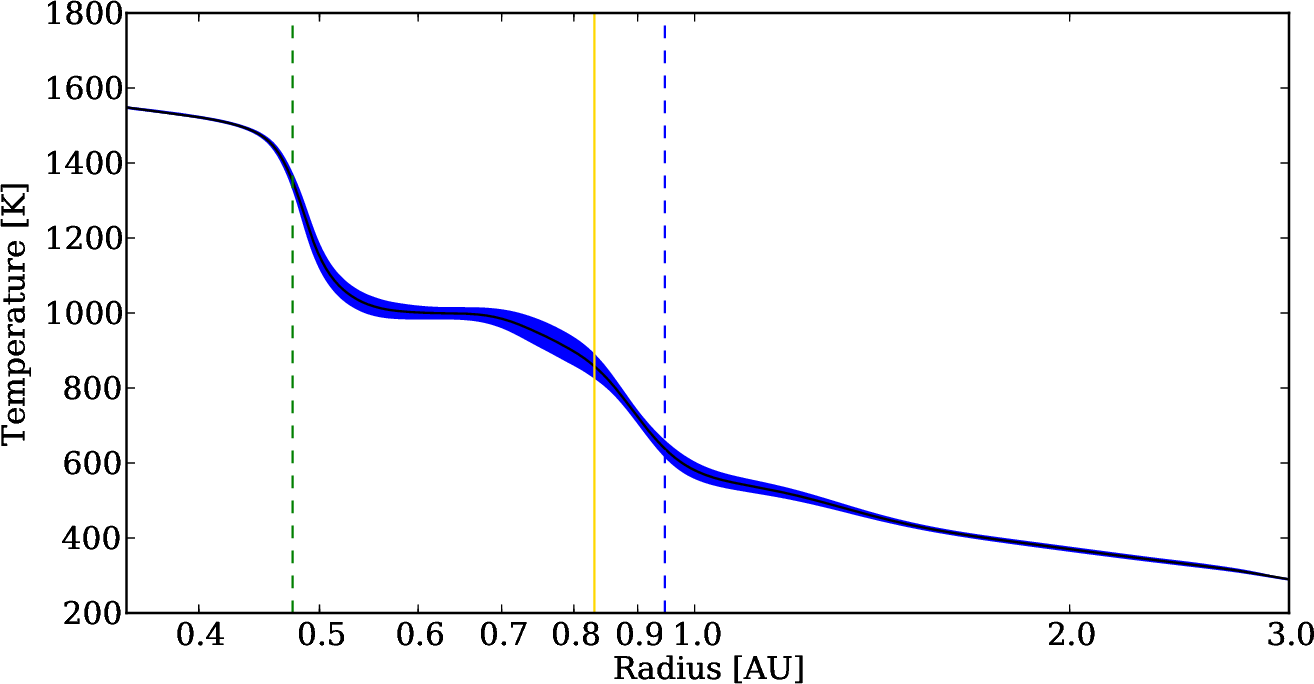}}
\caption{Time averaged and meridionally integrated radial profiles of the temperature for model \texttt{RMHD\_P0\_4}. Time average is done between 20 and 300 inner orbits. The blue shadow shows the standard deviations in the profile. Vertical lines show the midplane positions of the inner rim (green dashed line), the inner dead-zone edge (yellow line) and the dust concentration radius (blue dashed line).}
\label{fig:temp_av}
\end{figure}

We now take a deeper look at the time evolution of the disk structure. Fig.~\ref{fig:surftime} shows the time evolution of the surface density and three selected temperature contours. For the simulation runtime the disk remains stable. Overplotted temperature contours in steps of 400~K show the temperature variations over time at different disk locations. Close to the inner rim, the 1200~K contour show only small radial fluctuations. The 800~K and 400~K contours behind the inner rim move slightly radially inward by around 0.1 AU in the first 40 local orbits due to the extended height of the dust rim by magnetic fields, which causes a steeper temperature drop in the shadow behind the rim. 
Fig.~\ref{fig:temp_av} shows the time averaged radial temperature profile of model \texttt{RMHD\_P0\_4} overplotting the standard deviations. The temperature fluctuations remain small. The maximal deviations are around 10-30 K and appear due to variations of the rim surface and so the grazing angle variations of the irradiation.  

To summarize, the disk structure for model \texttt{RMHD\_P0\_4} remains stable over the simulation time. The model shows a quasi steady-state on dynamical timescales. However we emphasize that this model is not in inflow equilibrium on the longer accretion flow timescale. The initial surface density profile was calculated assuming $\alpha_{DZ}=10^{-3}$ for temperatures below 1000~K. In our 3D models we find values of $\alpha$ which are orders of magnitude smaller. In such a case the surface density should be much higher to balance the drop in accretion stress. In the discussion section we explore this point further.

\subsection{Time averaged results: Zero-net flux model}
\label{sec:timeav}

\begin{figure}
  \centering
  \resizebox{\hsize}{!}{\includegraphics{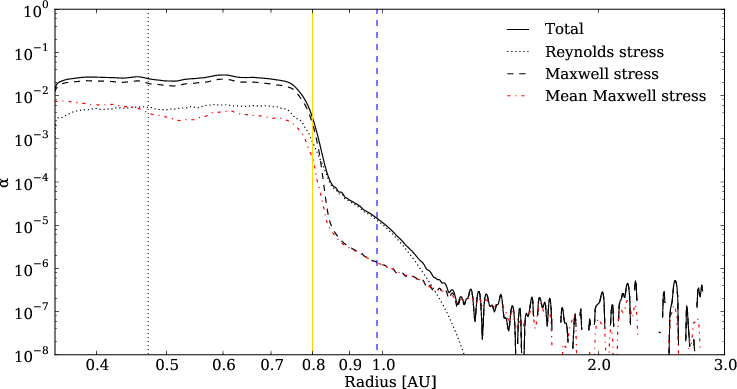}}
  \resizebox{\hsize}{!}{\includegraphics{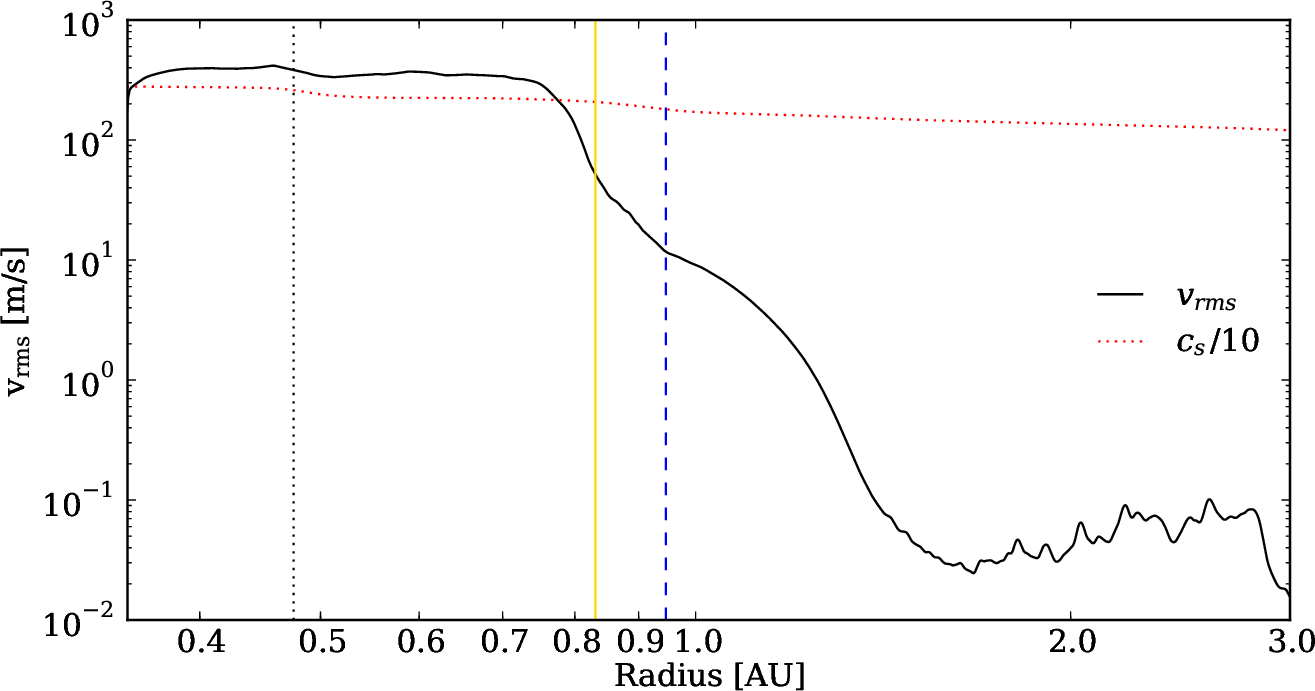}}
  \resizebox{\hsize}{!}{\includegraphics{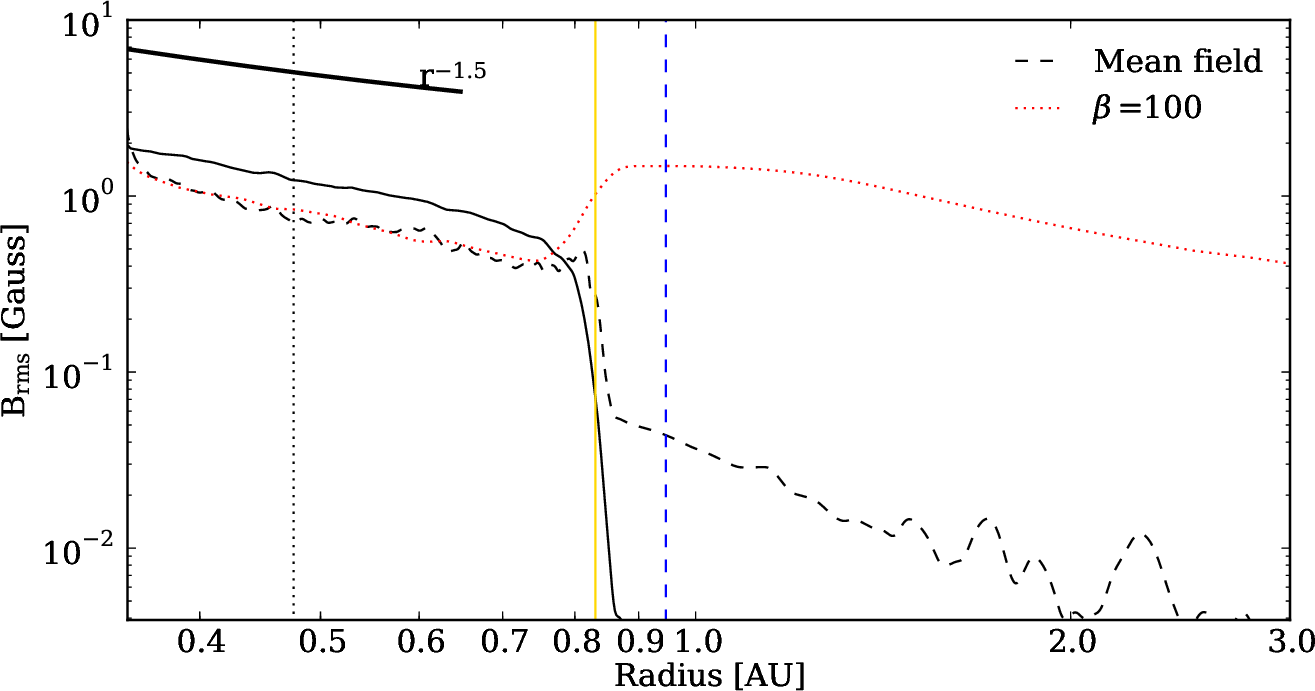}}
\caption{Time averaged and meridionally integrated radial profiles for model \texttt{RMHD\_P0\_4}, including the stress to pressure ratio $\alpha$ (top), the turbulent velocity (middle) and the turbulent magnetic field (bottom). Time average is done between 20 and 300 inner orbits. Vertical lines show the midplane positions of the inner rim (black dotted line), the inner dead-zone edge (yellow line) and the dust concentration radius (blue dashed line).}
\label{fig:turbq}
\end{figure}

The time averaged meridional distribution of the stress to pressure
ratio for model \texttt{RMHD\_P0\_4} is plotted in
Fig.~\ref{fig:al2D}, along with the positions of 
the inner rim and the dead-zone inner edge (green and yellow dashed
lines, respectively). The plot shows that the entire inner rim surface
is embedded in the turbulent region of the disk. The dead-zone edge is
located at $R \sim 0.8$ AU. Its shape is roughly vertical 
around the disk midplane ($|Z/R| < 0.12$) while in the upper disk
layers it curves radially outward, consistent with the fact that the
disk surface is radiatively heated by the central star. In agreement
with the results described above, turbulent activity quickly drops to
a very small value inside the dead-zone.

The time averaged and meridionally integrated 1D profile of the stress
to pressure ratio $\alpha$ is shown in Fig.~\ref{fig:turbq} (top
panel). The plot features a plateau with a constant value of
$\rm \alpha = 0.03$ over the entire inner disk, without any noticeable
change across the dust rim. In agreement with previously published
simulations performed in the ideal MHD limit
\citep{bra95,mil00,sim09I,dav10}, the Maxwell stress is around 3 times 
larger than the Reynolds stress. At the dead-zone inner edge, the
Maxwell stress drops sharply by roughly four orders of magnitude. By contrast, the Reynolds stress
decrease is more shallow, so that it dominates the total stress in a
small region between $0.8$ and $1.1$ AU which corresponds to roughly
$10$ disk scale heights. Over that region, the total stress takes
values that range between $\rm \alpha = 10^{-4}$ to $10^{-6}$ times the pressure. The
Maxwell stress amounts to about $10^{-6}$ and is governed by
mean magnetic fields which diffuse into the dead-zone. The mean Maxwell stress is computed from the volume averaged fields, see red dashed curve on the top panel. The relatively large Reynolds stress is
mainly due to density waves which are excited in the inner disk
turbulence \citep{hei09a}. Previous global MHD simulations of the
inner dead-zone edge have found similar amounts of Reynolds stress
close to the dead-zone inner edge \citep{dzy10,fau14}. These density waves
propagate into the dead-zone and are quickly damped after around
10 scale heights. In Appendix ~\ref{sec:damp} we review and discuss in
more detail the damping mechanism. Finally, outside of $1.1$ AU,
$\alpha$ drops further to values of about $10^{-7}$.   

The radial profile of the (mass weighted) turbulent velocity
fluctuations shown in Fig.~\ref{fig:turbq} (middle panel) displays similarities with the radial profile of $\alpha$. It plateaus inside $0.8$ AU, with values between $300$ and $400$ meters 
per second that correspond to several tens of percent of the sound
speed. At the rim position, there is a small decrease that we trace to the sudden temperature decrease and the associated drop in the sound speed (see red dotted line in
Fig.~\ref{fig:turbq}, middle panel). This is not surprising since the
value of $\alpha$ remains roughly constant across the inner rim. At
the dead-zone inner edge ($r \sim 0.8$ AU), the turbulent velocities
quickly drop by several orders of magnitude so that, at the position
of the dust concentration radius ($r \sim 1$ AU), the turbulent velocity fluctuations are reduced to values around $10$ m~s$^{-1}$.  

The radial profiles of the turbulent and mean magnetic field given in Fig.~\ref{fig:turbq} (bottom panel) were calculated by
performing a simple average over one scale height above and below the
midplane. Inside the MRI active region, the turbulent magnetic fields
show a power law radial profile with a slope of about $-3/2$ and a
strength of one Gauss at $0.6$ AU. This profile is slightly steeper than
the $r^{-1}$ radial profile found in toroidal net-flux global 3D
isothermal simulations \citep{flo11} and also recently again by
\citet{suz14} in global simulations with a vertical net flux
field. The radial profile of the turbulent and mean magnetic field
(the latter being dominated by the toroidal component) both follow the
profile that would be expected in a gas characterized by a constant
$\beta$ value, where $\beta=2 P/\bm{B}^2$ is the ratio between gas and
magnetic pressure (see red dotted line which corresponds to
$\beta=100$). At the dead-zone inner edge, the turbulent 
field strength quickly drops leaving a dominant mean magnetic field
inside the dead-zone. The plot shows that the mean magnetic field
diffuses outward in the dead-zone from its inner edge at $0.8$ AU. 

\begin{figure}
  \centering
  \resizebox{\hsize}{!}{\includegraphics{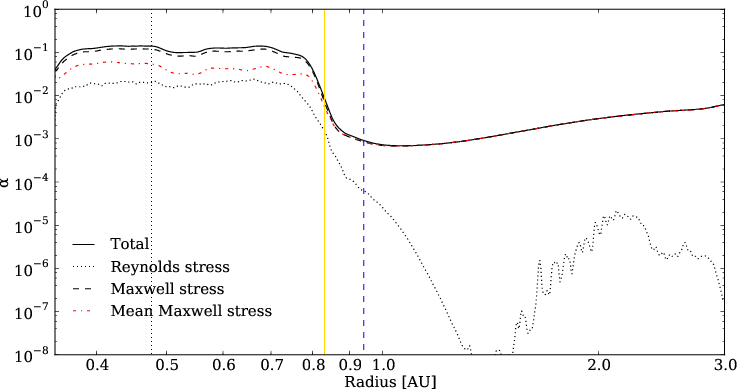}}
  \resizebox{\hsize}{!}{\includegraphics{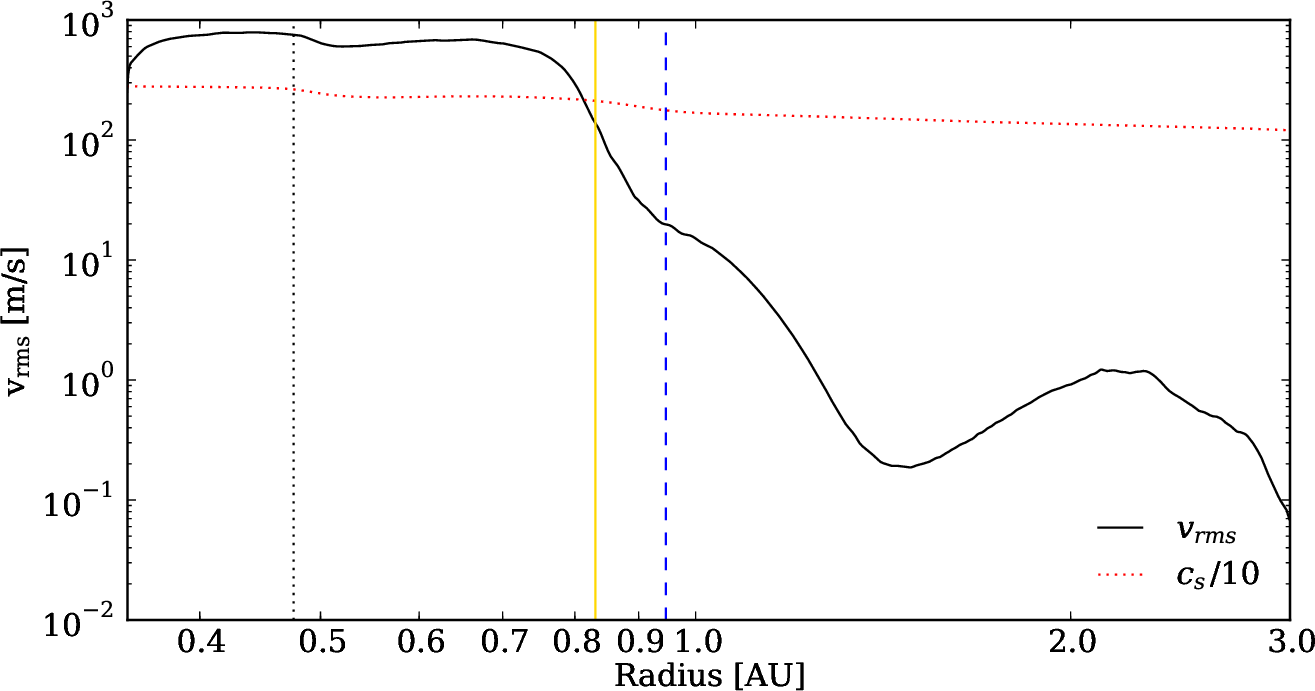}}
  \resizebox{\hsize}{!}{\includegraphics{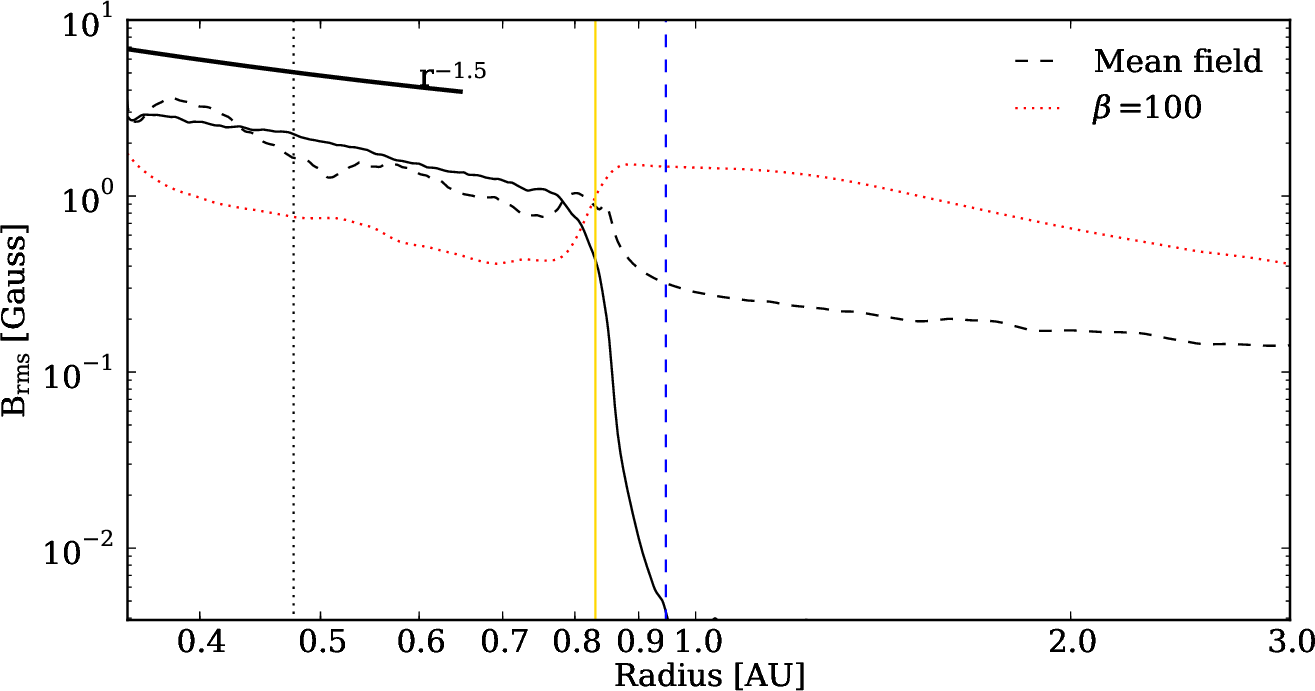}}
\caption{Time averaged and vertical integrated radial profiles for model \texttt{RMHD\_P0\_4\_BZ}, showing the stress to pressure ratio $\alpha$ (top), the turbulent velocity (middle) and the turbulent magnetic field (bottom). Time average is done between 20 and 70 inner orbits. Vertical lines show midplane positions of the inner rim (black dotted line), the inner dead-zone edge (yellow line) and the dust concentration radius (blue dashed line).}
\label{fig:turbq_vert}
\end{figure}

\begin{figure*}
\begin{minipage}{0.49\textwidth}
  \resizebox{9cm}{!}{\includegraphics{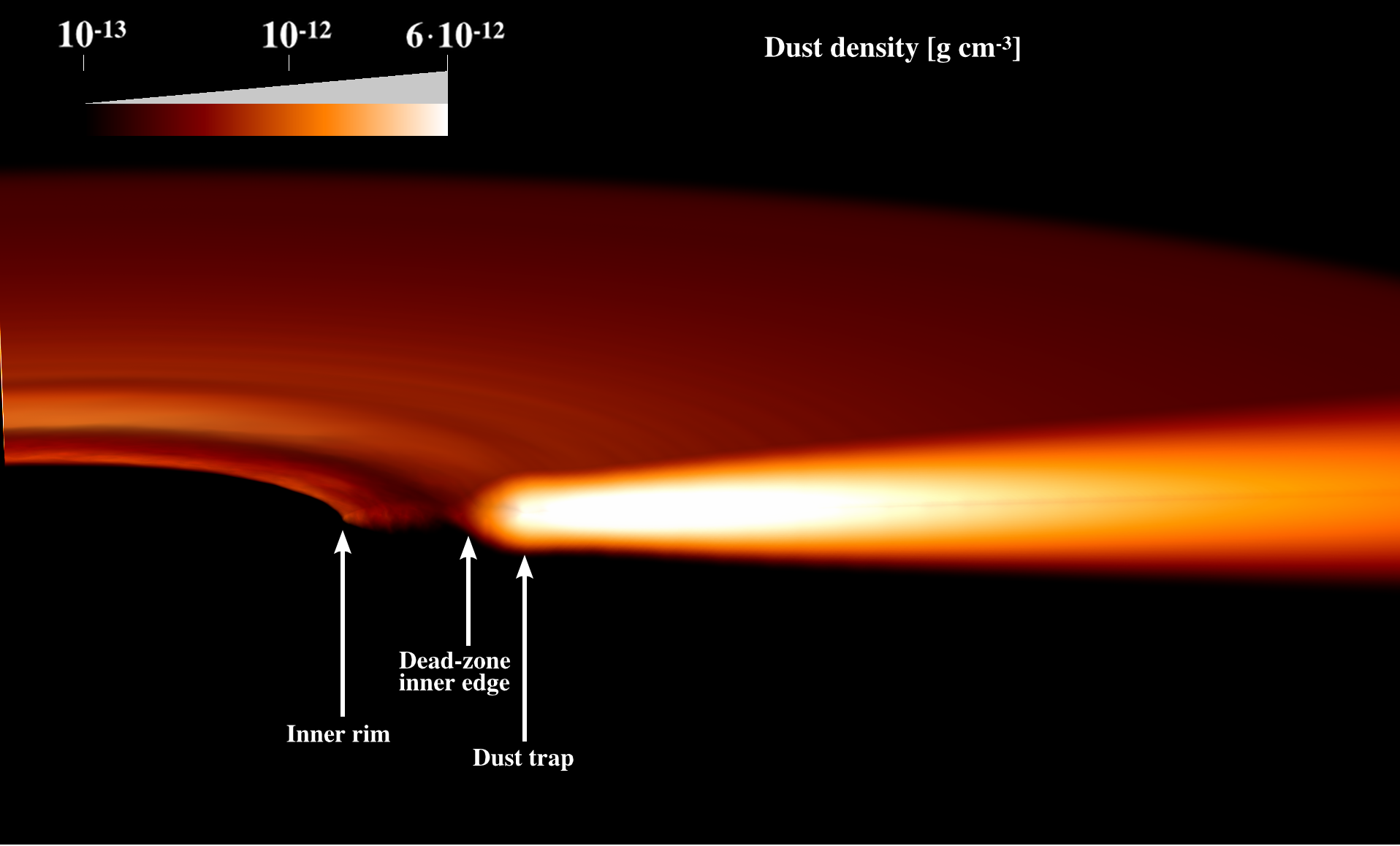}}
\end{minipage}
\begin{minipage}{0.49\textwidth}
  \resizebox{9cm}{!}{\includegraphics{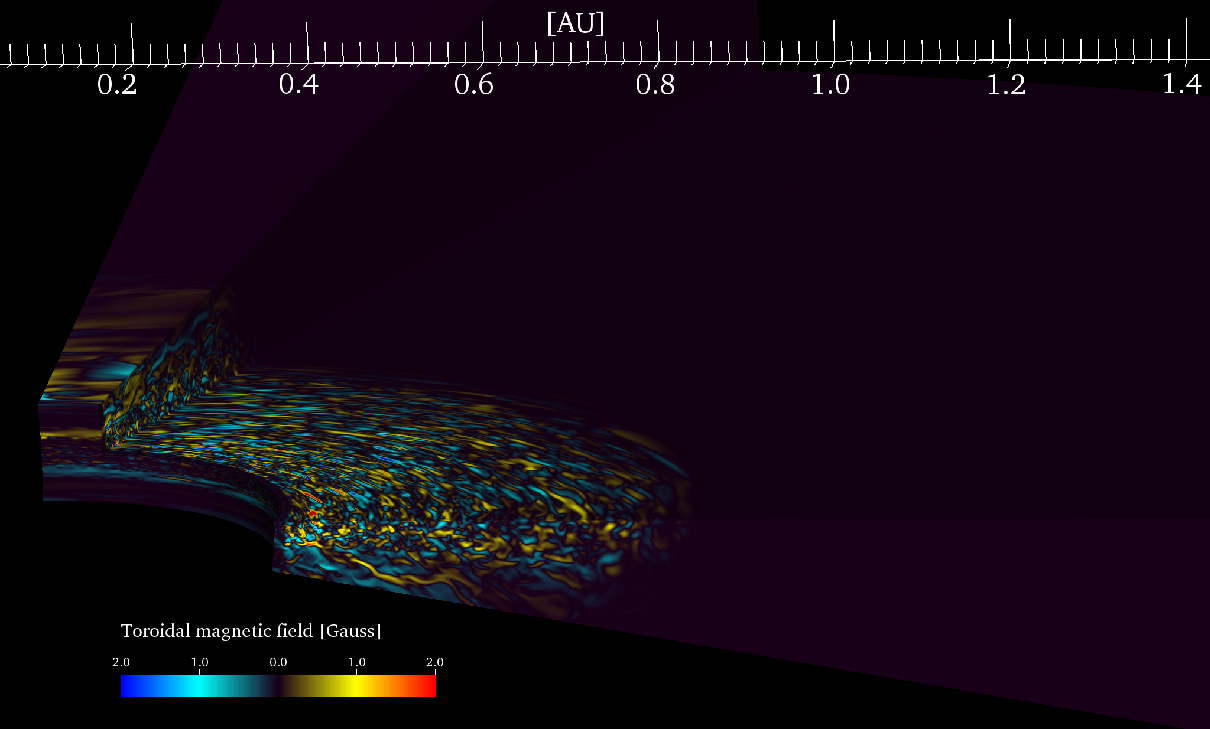}}
\end{minipage}
\caption{3D volume rendering of the dust density (left) and the toroidal magnetic field (right) for model \texttt{RMHD\_P1\_6} after 50 inner orbits. On the left, the location of the inner rim (0.47~AU), the dead-zone edge (0.8~AU) and the dust concentration radius (1~AU) are annotated. On the right, the domain's top half is cut away to show the magnetic field on the midplane.}
\label{fig:3d}
\end{figure*}

\subsection{Stronger turbulence: the net flux model}
\label{sec:vert}

In this section we present the results of model
\texttt{RMHD\_P0\_4\_BZ}. It is intended to simulate the conditions in a protoplanetary disk which is under the influence of a vertical magnetic field. Model \texttt{RMHD\_P0\_4\_BZ} was computed by restarting model
\texttt{RMHD\_P0\_4} after $210$ inner orbits, adding a uniform vertical magnetic field (see Appendix~\ref{ap_1} for
details) whose strength is such that $\rm \beta=3.5 \times 10^{4}$ at
$1$ AU in the disk midplane. We find that the radial locations of the
inner rim, the dead-zone edge and the dust concentration are only
weakly modified compared to model \texttt{RMHD\_P0\_4}. 

Fig.~\ref{fig:turbq_vert} (top panel) plots the time averaged and
meridionally integrated radial profile of the stress to pressure
ratio. Time averaging is done between 20 and 70 inner orbits. The model
\texttt{RMHD\_P0\_4\_BZ} shows a substantially increased turbulent
activity compared to model \texttt{RMHD\_P0\_4}, with a high plateau
of $\alpha \sim 0.1$ in the inner disk. Such high $\alpha$ values are
expected in the ideal MHD limit when the disk is threaded by a
vertical net flux magnetic field \citep{bai13}. The mean Maxwell
stress is large even in the MRI active region and accounts for about
half of the angular momentum transport. It drops by one order of
magnitude only at the dead-zone inner edge (see red dash-dotted line)
and dominates the total stress in the bulk of the dead-zone, with a
typical $\alpha$ of order $10^{-3}$. Such large values generated by
axisymmetric magnetic fields have also been found in local box
simulations that include a Ohmic dead-zone only (see model V1 by
\citet{tur07}, model X1d by \citet{oku11} or model D1-NVFb by
\citet{gre12}).

Next the middle panel of Fig.~\ref{fig:turbq_vert} shows the radial
profile of the time averaged and mass weighted turbulent velocity
fluctuations. Compared to \texttt{RMHD\_P0\_4}, $v_{rms}$ is increased
by a factor of two to three in the inner disk. Finally, the time
averaged turbulent and mean magnetic fields are shown in
Fig.~\ref{fig:turbq_vert} (bottom panel). Again, in the MRI active
region, we find an increase by a factor two to three compared to model
\texttt{RMHD\_P0\_4}. Their radial slopes, however, remain similar in
this region. The biggest difference is the presence of a large mean
field in the whole domain (compare black dashed lines in the bottom
panel of Fig. \ref{fig:turbq} and Fig. \ref{fig:turbq_vert}). This is
particularly true in the dead-zone, where the mean field in model
\texttt{RMHD\_P0\_4\_BZ} is almost two orders of magnitudes larger
than the typical value we found in model
\texttt{RMHD\_P0\_4}. Finally, we caution that the total duration of the simulation for model \texttt{RMHD\_P0\_4\_BZ} remains fairly small. We refer the reader to the discussion section for
more details.  

\begin{figure}
  \centering
  \resizebox{\hsize}{!}{\includegraphics{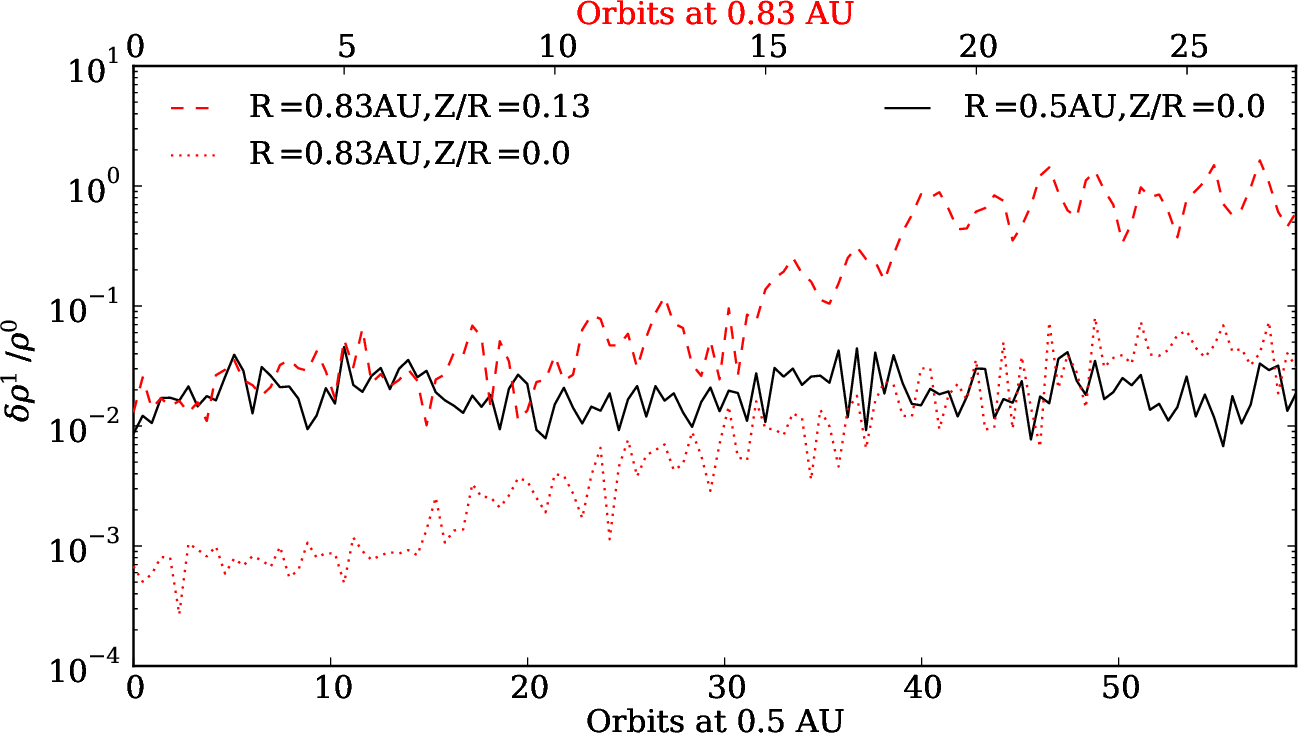}}
\caption{Azimuthal density variations in the largest azimuthal mode over time in model \texttt{RMHD\_P1\_6} at three different location inside the disk. After 20 local orbits (see upper x-axis), the non-axisymmetric perturbation close to the dead-zone inner edge grows by two order of magnitude, reaching order unity in the upper layers (red dashed line).}
\label{fig:nonaxis}
\end{figure}
\begin{figure}
\centering
%  \hspace{-1.3cm}
%\begin{minipage}{0.4\textwidth}
\resizebox{\hsize}{!}{\includegraphics{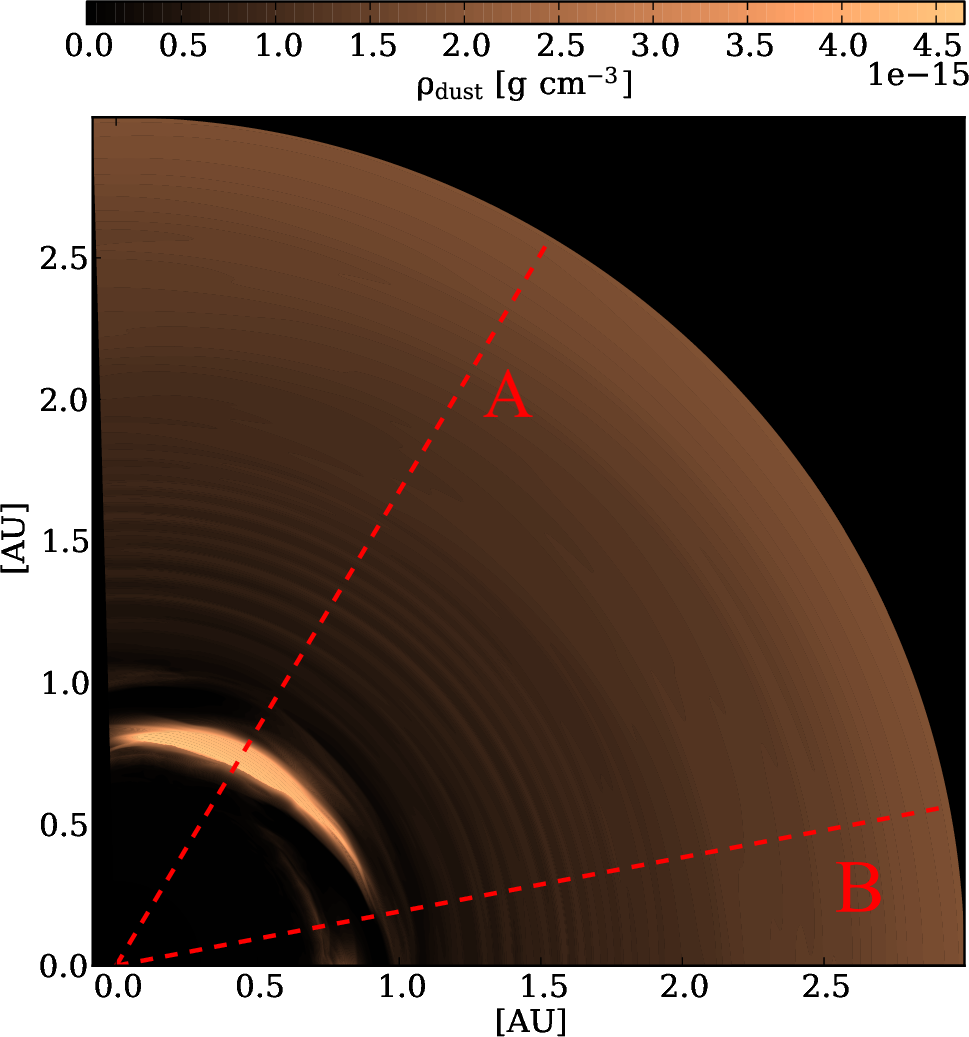}}
\resizebox{\hsize}{!}{\includegraphics{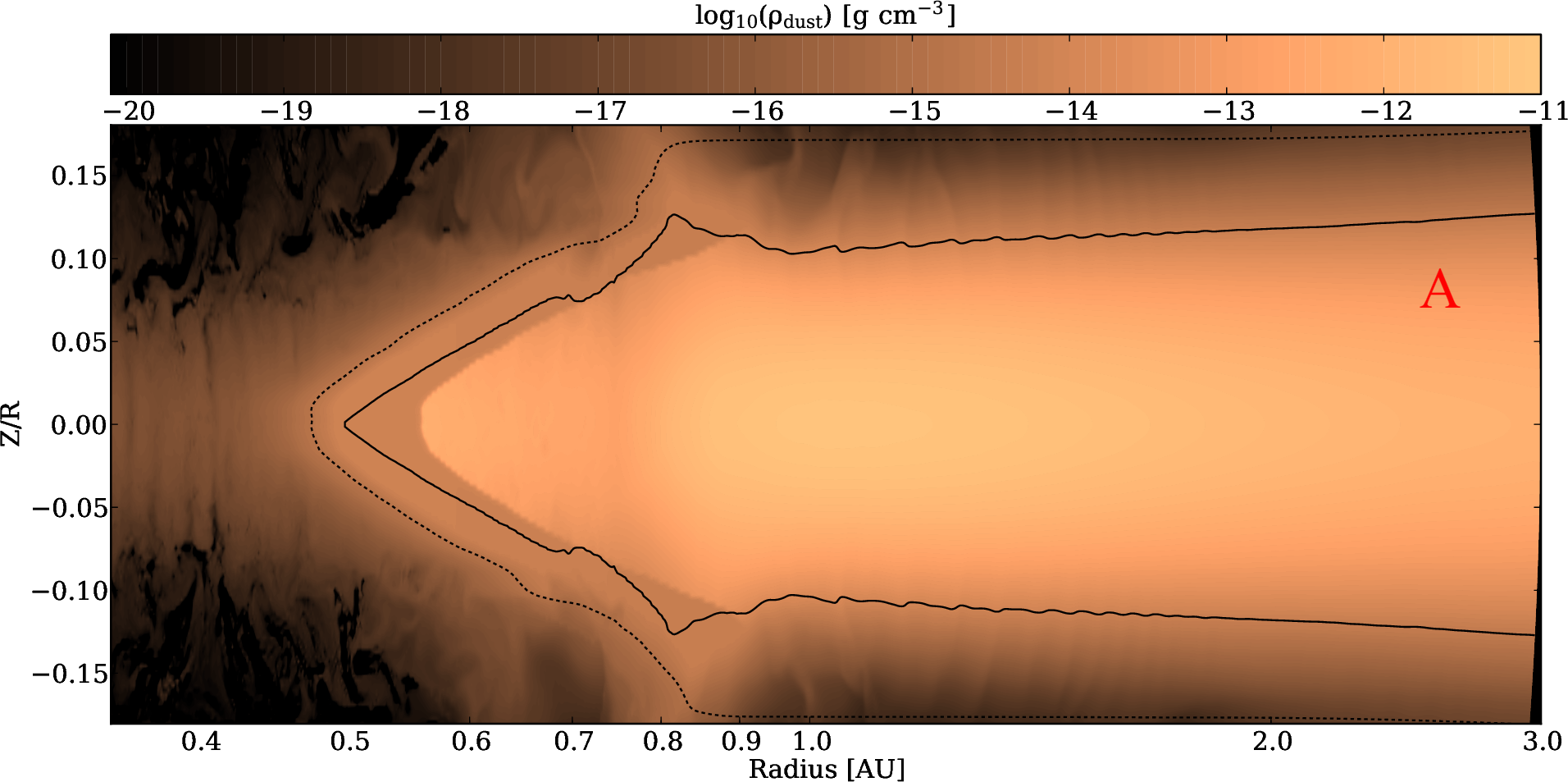}}
\resizebox{\hsize}{!}{\includegraphics{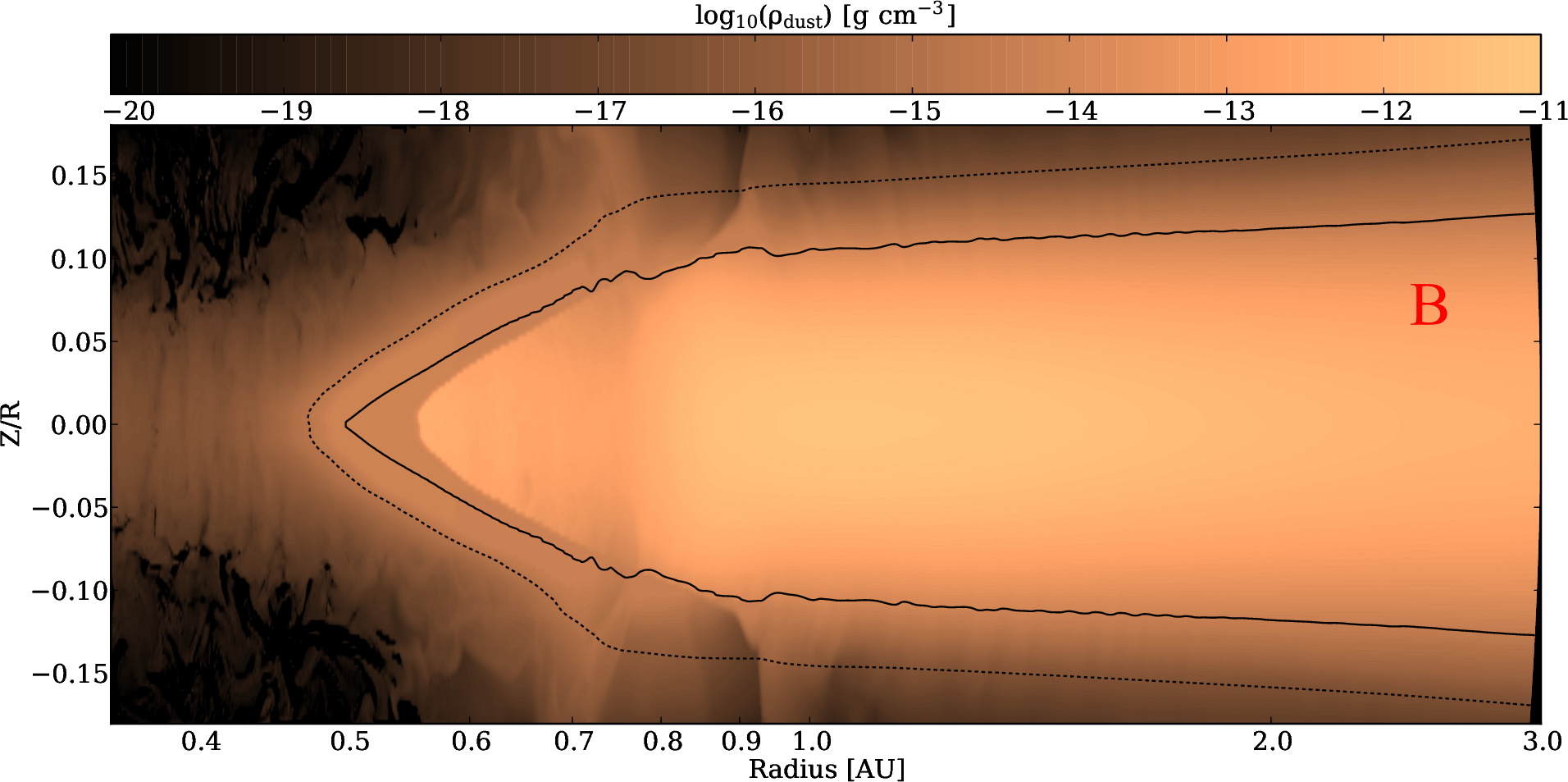}}
\caption{Cross-sections through the dust density in different orientations. The face-on view (top) is cut at the height Z/R=0.13. The red dashed line shows the positions of the two slices with the vortex (middle) and without (bottom). The black lines indicate optical depth unity for the irradiation (dotted line) and for the thermal emission (solid line).}
\label{fig:dustsl}
\end{figure}
\subsection{Long lasting non-axisymmetric perturbations}
\label{sec:large}

In this section, we investigate the potential growth of non-axisymmetric
structures. This is done by using
model \texttt{RMHD\_P1\_6}, which is similar to model
\texttt{RMHD\_P0\_4} but features a larger azimuthal extent. The
initial conditions for this simulation are generated using 
a snapshot of the flow in model \texttt{RMHD\_P0\_4} after 150 inner
orbits and periodically repeating the azimuthal domain four
times. Velocity perturbations of the order $\rm 10^{-4} c_s$ are
applied cellwise to each component to break the symmetry. Model
\texttt{RMHD\_P1\_6} quickly reaches a new turbulent state, albeit
with statistical properties similar to model \texttt{RMHD\_P0\_4} (see
Appendix~\ref{app:comp}). The 3D rendering of the dust density after
50 inner orbits is shown in Fig.~\ref{fig:3d} (left panel). The plot
confirms that dust particles are found between the rim and the dead zone
inner edge in a highly turbulent environment. As discussed in the
previous sections, the dust density increases sharply at the dead zone inner
edge, following a similar increase in the gas surface density that is
due to the drop of the accretion stress \citep{flo16}. The right panel
shows the tangled structure of the turbulent magnetic field. It is
dominated by the toroidal component which reaches amplitudes of
several Gauss.

We next compute discrete Fourier transforms of the density along
azimuth. We focus on three different locations: (1) the midplane at
0.5 AU, which is fully turbulent, (2) the upper layer close to the inner
rim surface and the dead-zone edge (R=0.83 AU and Z/R = 0.13), and (3) the
midplane at 0.83 AU. The results are summarized in
Fig.~\ref{fig:nonaxis}. Initially, the large scale density variations
are weak. They amount to a few percent in both the turbulent midplane 
at $R=0.5$ AU and in the disk upper layers at $R=0.83$ AU (see black
solid and red dashed lines). They are even smaller in the disk midplane
at $R=0.83$ AU (i.e. at the dead-zone edge) where they only reach value
of $\sim 10^{-3}$ (red dotted line). In the MRI active region, these 
density perturbations do not grow for $60$ local orbital periods ($\sim$
129 inner orbits). However, there is a clear increase by about two
orders of magnitude at the location of the dead-zone edge (both
red lines). In particular, the relative perturbations reach order unity in the disk upper layers at $R=0.83$ AU. The presence of
a sharp surface density change at that location and the growth
timescale of $\sim 20$ local orbits both suggest the
Rossby wave instability (RWI) \citep{lov99,meh13}. This is
confirmed by the appearance of a localized vortex (not shown)
characterized by a midplane relative vorticity of about $(\nabla
\times v)_z/\Omega \approx -0.3$ in the vortex core. Similar values
have been reported in the literature for vortices produced by the
RWI \citep{lyr12,meh13,flo15}. Fig.~\ref{fig:nonaxis} indicates that the vortex is not growing anymore after 20 local orbits. Regarding the expected lifetime of the vortex we can only make predictions based on its shape. The vortex has an extent of $\sim 2$H in radius and around $\sim 22$H in azimuth ($\rm H/R \sim 0.04$ at the vortex
location). Such an elongated vortex is known to be vulnerable to
elliptical instabilities \citep{les09}, which will limit its subsequent growth. An alternating vortex growth and destruction, as it was found in \citep{flo15}, could be also possible.

The vortex appears in Fig.~\ref{fig:dustsl} (top
panel) in the snapshot of the dust density after 150
inner orbits as a clear non-axisymmetric density maximum. The middle and
bottom panels of Fig.~\ref{fig:dustsl} show the dust density in the
$R$-$Z/R$ plane inside and outside the vortex (along the lines labeled
``A'' and ``B'' on the top panel, respectively). Inside the overdensity,
the irradiation optical depth unity line (or, equivalently, the height
of the rim) is increased vertically (see the dotted lines on the
middle and bottom panels of Fig.~\ref{fig:dustsl}): at $R=0.85$ AU, we
measured $(Z/R)_{\tau_*=1}=0.17$ for cut ``A'', and
$(Z/R)_{\tau_*=1}=0.14$ for cut ``B''. As will be further discussed in
section~\ref{sec:var}, such a difference in height is enough to
create an extended shadow on the disk beyond. 

\begin{figure}
  \centering
  \resizebox{\hsize}{!}{\includegraphics{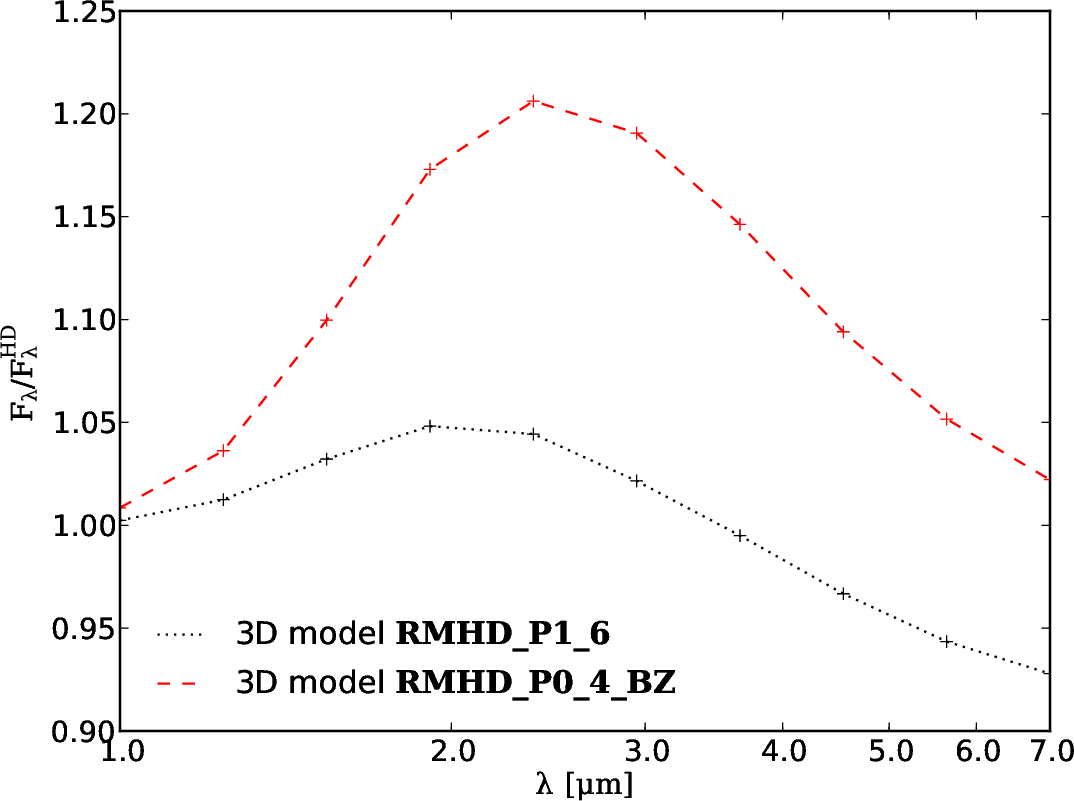}}
\caption{SED calculated for the 3D radiative MHD models \texttt{RMHD\_P1\_6} (black dotted line) and model \texttt{RMHD\_P0\_4\_BZ} (red dashed line), normalized over the SED from the initial 2D radiation HD model.}
\label{fig:sed}
\end{figure}
 \begin{figure}
  \centering
  \resizebox{\hsize}{!}{\includegraphics{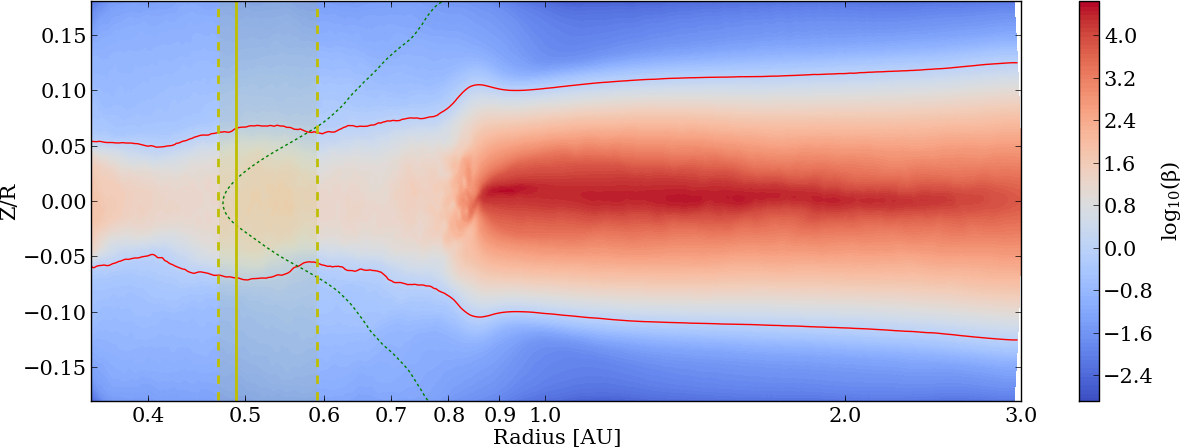}}
\caption{Space (azimuthally) and time (50 inner orbits) averaged plasma $\beta$ profile in the $R$-$Z/R$ plane for model \texttt{RMHD\_P0\_4\_BZ}. The line of plasma $\beta$ unity (red solid line) and the $\tau_*=1$ line (green dashed line) are overplotted. The solid yellow line shows the position of the peak emission at $\rm 2.2 \mu m$ assuming a face-on orientation. The yellow dotted lines and the shading demonstrates the spatial extent where 50\% of the total emission at $\rm 2.2 \mu m$ is coming from. }
\label{fig:plasma}
\end{figure}
\begin{figure}
  \centering
  \resizebox{\hsize}{!}{\includegraphics{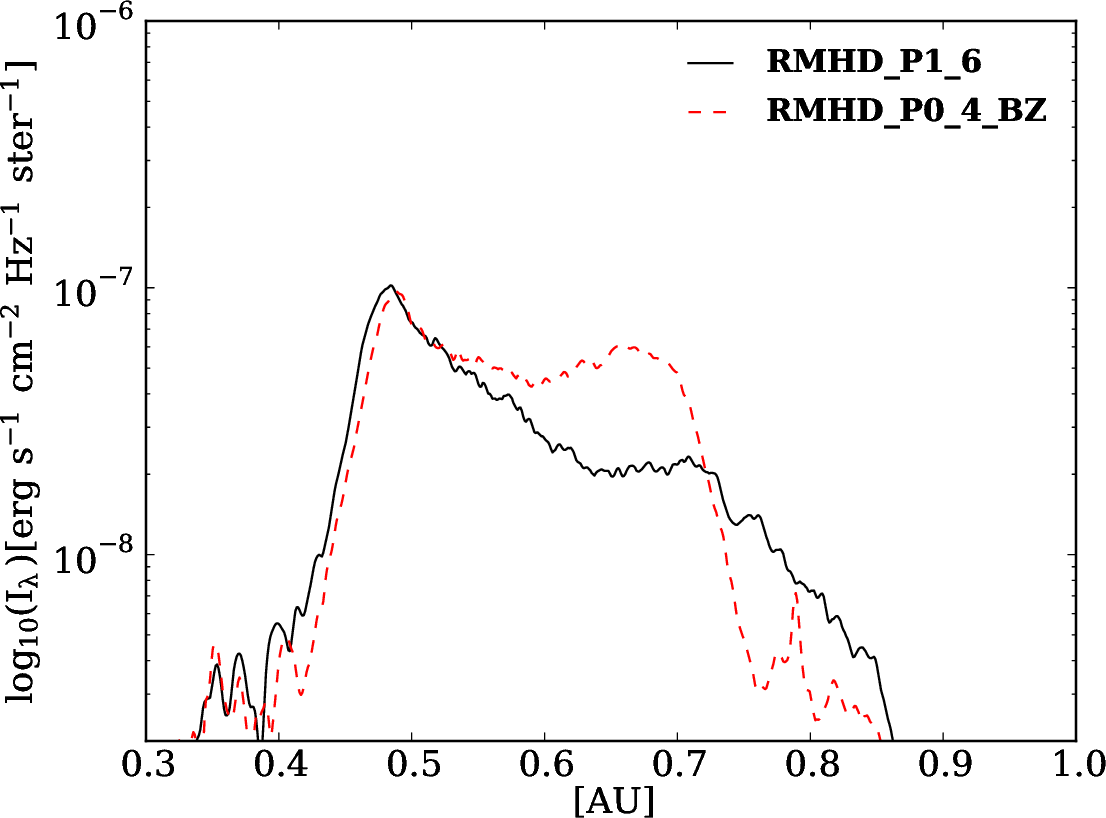}}
\caption{Radial cut from a face-on synthetic image calculated at $2.2 \mu m$ for the model \texttt{RMHD\_P1\_6} and \texttt{RMHD\_P0\_4\_BZ}. }
\label{fig:imcut}
\end{figure}
\section{Observational constraints}

In this section we post-process our models with Monte-Carlo
radiation transfer tools in order to translate the results described
in the previous sections into observational constraints. This is done using the Monte Carlo radiative transfer code
RADMC3D \citep{dul12} for which we use the same parameters as in \citet{flo16}. For more details on the
post-processing and the RADMC configuration we use, we refer the reader to Appendix~\ref{sec:ap3}.
We first focus on the disk spectral energy distribution (SED) in
Section~\ref{sec:sed} and next compute synthetic images in
Section~\ref{sec:syn}. Then, we discuss the time variability
associated with the disk dynamics in Section~\ref{sec:var}.

\subsection{SED}
\label{sec:sed}

First, we determine the SED of the models \texttt{RMHD\_P1\_6} and
\texttt{RMHD\_P0\_4\_BZ} and compare them with the initial radiation
HD models. The system is seen inclined 45$\rm ^\circ$ from face-on and the
azimuthal domain is repeated to cover the entire $2 \pi$ azimuthal
range. We calculate the flux at seven individual wavelengths between 1
and 7 $\rm \mu m$ as it is the relevant wavelength range for our
domain size and temperature range. 
The results are plotted in Fig.~\ref{fig:sed}. The 3D
radiation non-ideal MHD model \texttt{RMHD\_P1\_6} is very close to
the 2D radiation hydrodynamical model, with a modest increase of the
emission around $\rm 2 \mu m$ of $5 \%$ due to the weak magnetized corona. By contrast, model
\texttt{RMHD\_P0\_4\_BZ}, for which the magnetic activity is much
stronger, shows a significant increase of about $20 \%$ compared to
the hydrodynamical model.  

A discussion of the origin of the emission arising at different
wavelengths is enlightening to understand this difference. Most of the
NIR emission is thermal in origin and comes from the surface located where Z/R $<$ 0.05 and R $<$
0.6 AU. At these locations, the magnetic field is weak and does not
alter the density distribution: the area of the emitting region
is unchanged. This can be illustrated with the help of the quantity
$\beta$, which indicates the importance of gas relative to magnetic pressure. 
Fig.~\ref{fig:plasma} shows the distribution of $\beta$ in
the $R$-$Z/R$ plane for model \texttt{RMHD\_P0\_4\_BZ}, averaged in
time and in azimuth. The upper layers are magnetically dominated
($\beta \leq 1$) while the midplane region remains gas pressure 
dominated ($\beta \geq 1$). The equipartition line ($\beta=1$) stays
above the rim surface ($\tau_* =1$) for $\rm R< 0.6\, AU$. Most
of the J and K band NIR emission is coming from a narrow region: the
solid yellow line in Fig.~\ref{fig:plasma} marks the location of the peak
emission (solid) and the area between the dashed yellow lines
corresponds to the location where 50\% of total flux at $\rm 2.2 \mu
m$ is emitted (at face-on orientation). Although it is shown
here for model \texttt{RMHD\_P0\_4\_BZ}, this result is similar 
for model \texttt{RMHD\_P1\_6}. Fig.~\ref{fig:imcut} helps to understand
the increased NIR emission in model \texttt{RMHD\_P0\_4\_BZ}: it shows
a radial cut from the synthetic image at $\rm 2.2 \mu m$ for both
models. The regions outward from $0.6$ AU mainly contribute to the
increased emission in model \texttt{RMHD\_P0\_4\_BZ}. As seen in
Fig.~\ref{fig:plasma}, the magnetic pressure is stronger than the gas pressure at the position of 
$\tau_* =1$ which results in increased magnetic support, and
thereby a shallower density profile in the disk upper layers. By
contrast, model \texttt{RMHD\_P1\_6} presents weaker magnetic
activity and a thinner magnetically supported corona which leads to
the SED profile being closer to the hydrodynamical model. Overall,
these results suggest that magnetic field are able to increase the
emission between $2$ to $3$ $\rm \mu m$ by $5$ to $20\%$. We discuss
the possibility of obtaining an even higher NIR excess in
section~\ref{sec:discussion}. 

\begin{figure}
  \resizebox{\hsize}{!}{\includegraphics{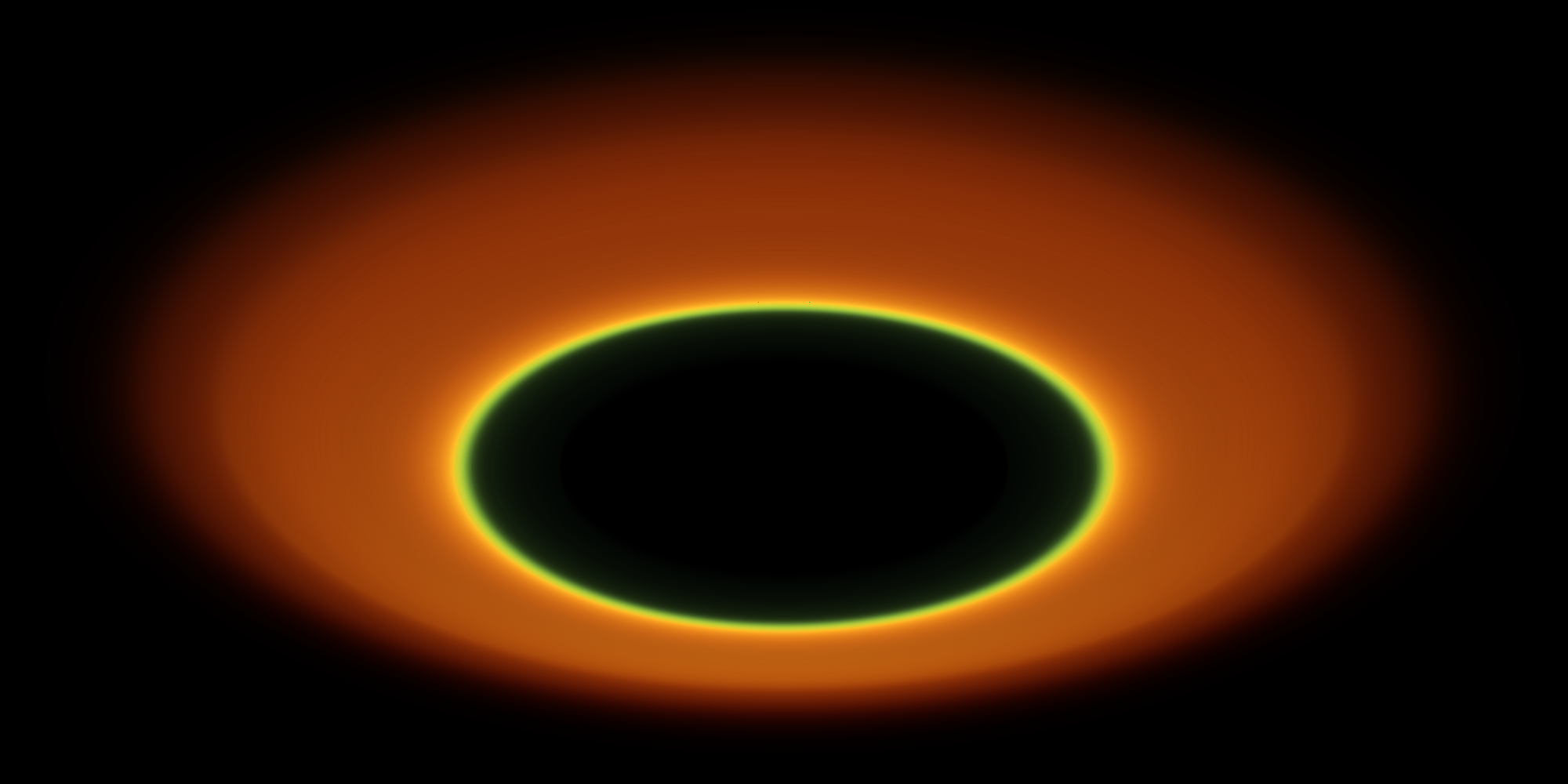}}
  \resizebox{\hsize}{!}{\includegraphics{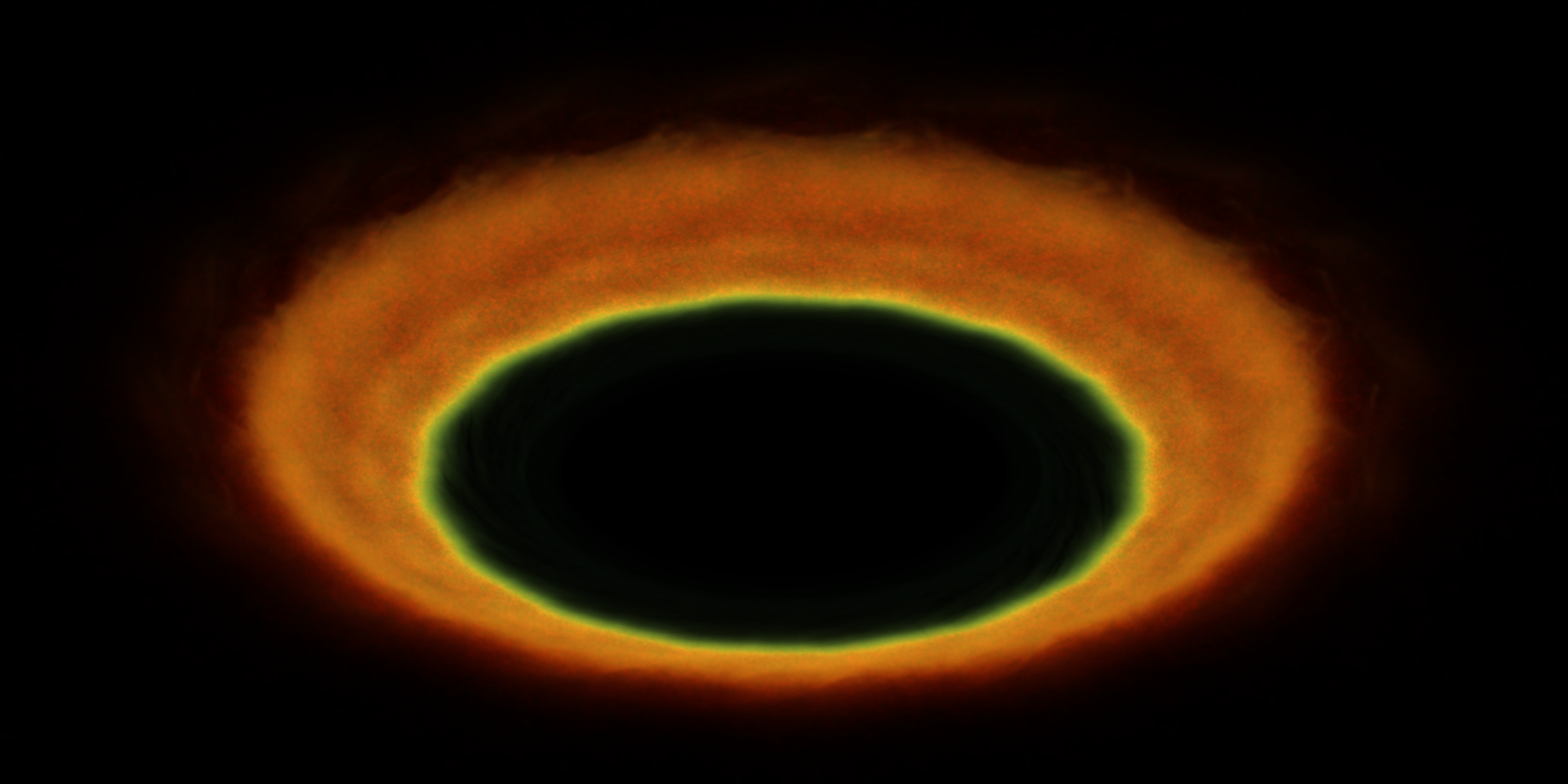}}
  \resizebox{\hsize}{!}{\includegraphics{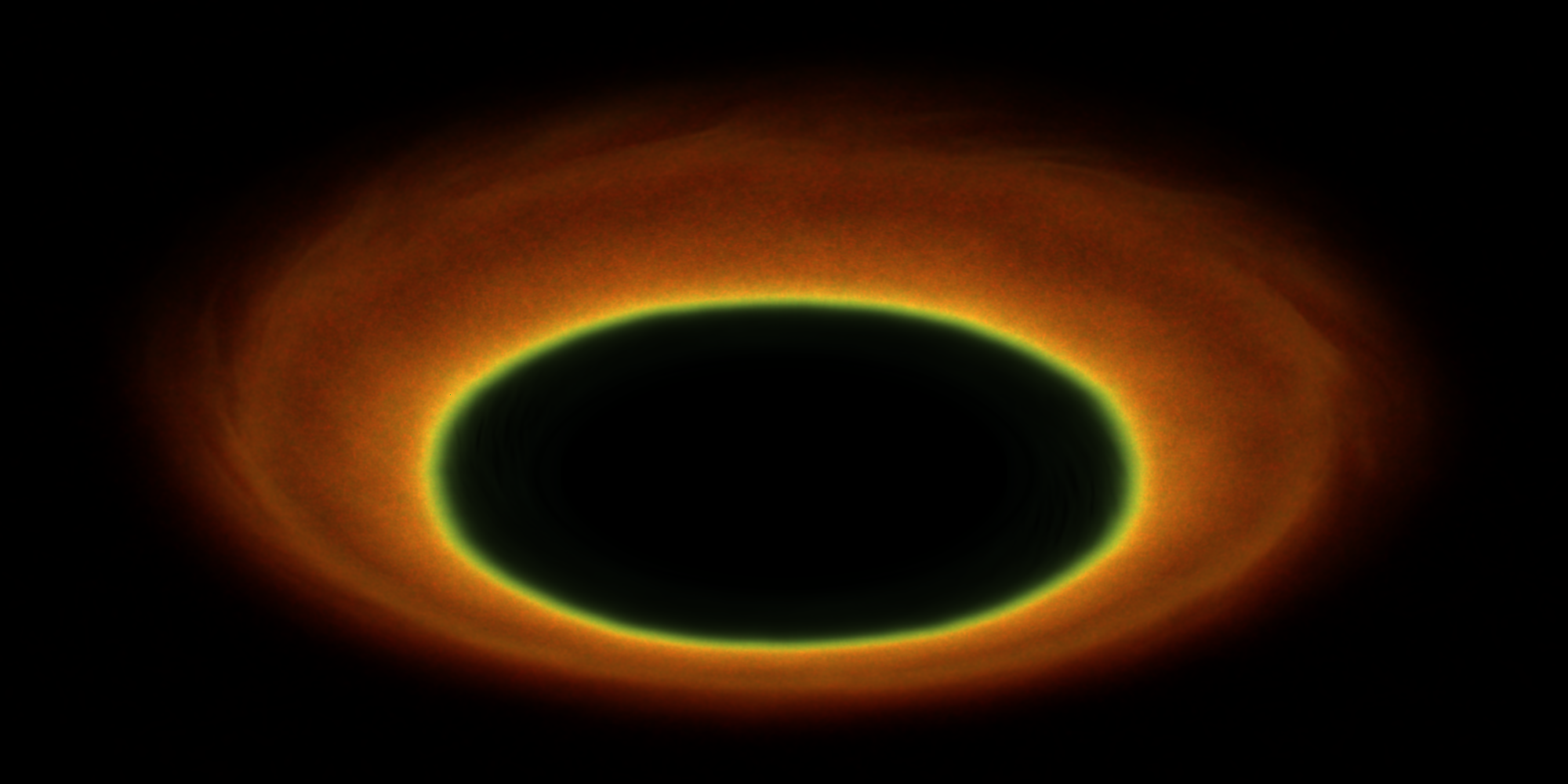}}
  \resizebox{\hsize}{!}{\includegraphics{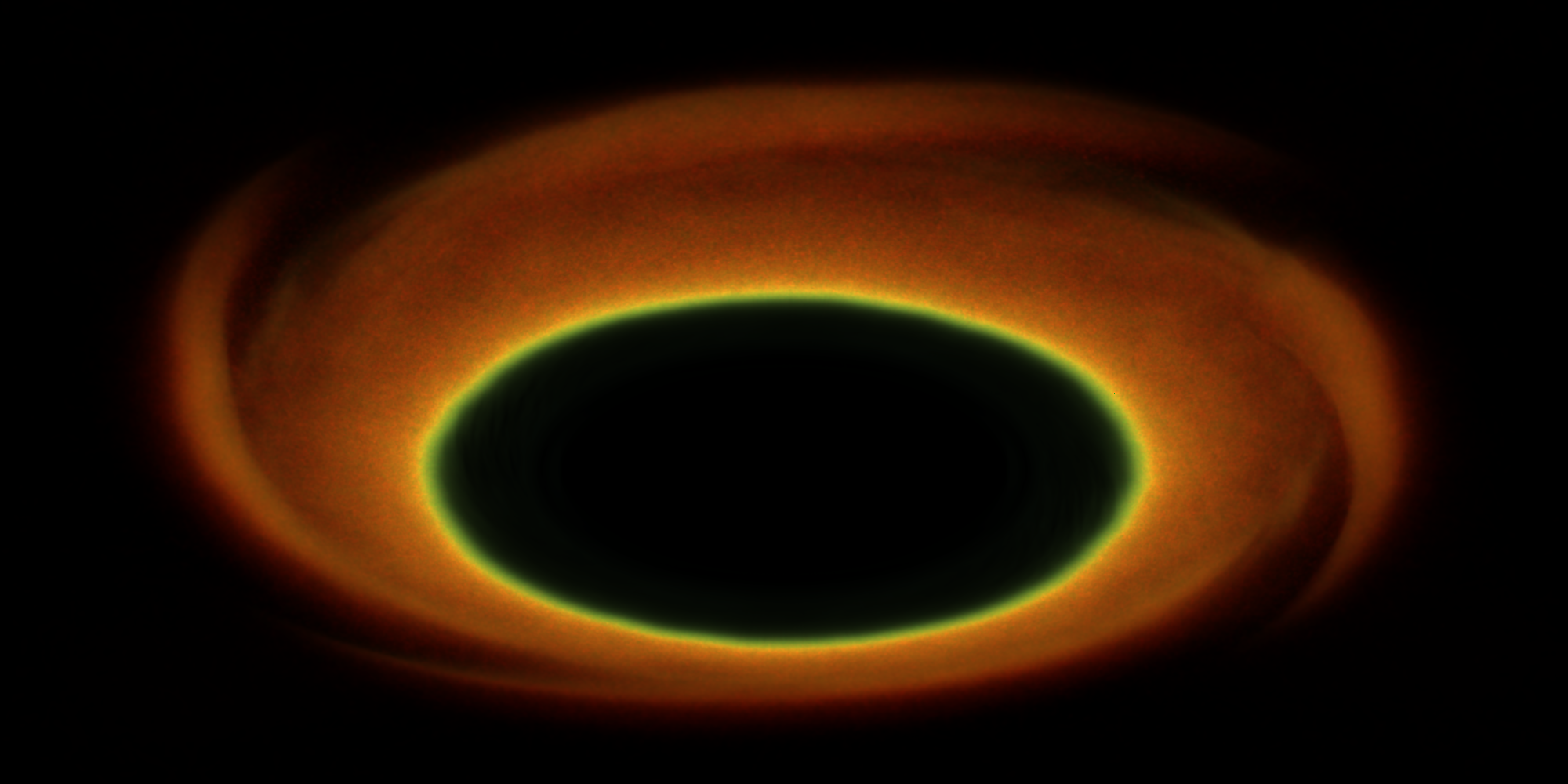}}
\caption{Synthetic images of the initial radiation HD model (top) and the global 3D radiation non-ideal MHD models \texttt{RMHD\_P0\_4\_BZ} (second) after 50 inner orbits, and model \texttt{RMHD\_P1\_6} (third and bottom) after 50 and 150 inner orbits. The field of view is 2 AU wide, and the system is inclined 60$\rm ^\circ$ from face-on. The blue, green, and red channels in each image correspond to wavelengths 1.25, 2.2, and 4.8 $\rm \mu$m, corresponding to J, K, and M bands, respectively. The plot shows linear intensity. Normalization is done over the maximum intensity at 4.8 $\rm \mu$m. We note that 2$\pi$ coverage is obtained by repetitively extending the azimuthal domain, which leads to the repeating spiral structure for model \texttt{RMHD\_P1\_6} (bottom).} 
\label{fig:syn_sed}
\end{figure}

\subsection{Synthetic images}
\label{sec:syn}

Synthetic images are shown in Fig.~\ref{fig:syn_sed} corresponding to the radiation hydrodynamical
model (top panel) and the global 3D radiation non-ideal MHD models
(bottom three panels). For all cases, the images cover a region which
is approximately $\rm 2 \, AU$ wide. The fluxes at $1.25$, $2.2$, and $4.8$ $\rm
\mu$m from the blue, green and red channels on a shared linear color scale normalized over the maximum intensity at 4.8 $\rm \mu$m. In the synthetic images of the 3D models computed at early
times during the simulations (second and third panels), small turbulent
structures can be identified, especially in the uppermost layers of the
near-infrared emitting region. Model \texttt{RMHD\_P0\_4\_BZ} 
(second panel) shows a slightly narrower and brighter ring compared
to model \texttt{RMHD\_P1\_6} (third panel). This is because the rim 
surface is slightly steeper compared to the other models (see also
the $\tau_*=1$ line in Fig.~\ref{fig:plasma}) owing to the magnetic support. At later time during
the evolution of model \texttt{RMHD\_P1\_6}, the effect of the vortex
becomes visible in the synthetic image of the NIR emission 
(bottom panel in Fig.~\ref{fig:syn_sed}) as a spiral pattern
visible in the M band emission. We note that the $m=4$ pattern comes from duplicating the simulation domain 
in the azimuthal direction before viewing the snapshot. Global $2\pi$ hydrodynamical calculations indicate such multiple RWI vortices merge leaving a single $m=1$ pattern \citep{lyr12,meh12}.

%% In our previous models by
%% \citet{flo16}, we have shown the dominant emission from the 2.2, and
%% 4.8 $\rm \mu$m compared to the shorter wavelength. 

\begin{figure}
\resizebox{\hsize}{!}{\includegraphics{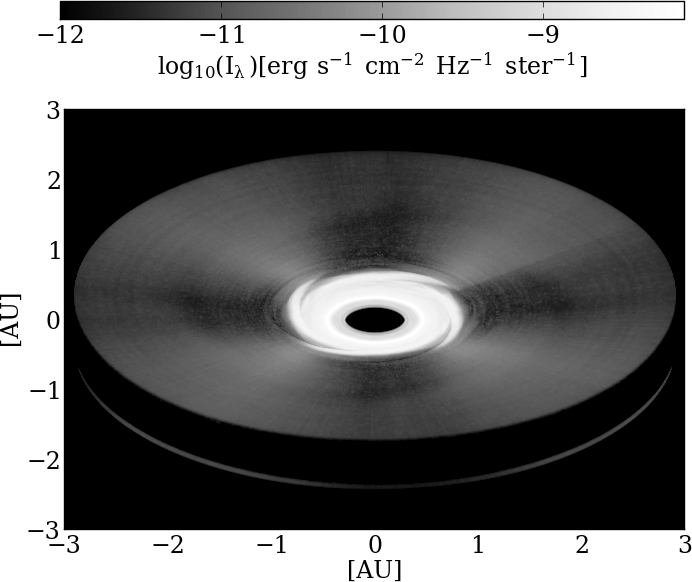}}
\caption{Synthetic image at $\rm 0.3\,  \mu m$ after 150 inner orbits for model \texttt{RMHD\_P1\_6}. The system is inclined by 45$\rm ^\circ$. We note that the 2$\pi$ coverage is obtained by repetitively extending the azimuthal domain, which leads to the m=4 pattern.} 
\label{fig:varscat}
\end{figure}

Using the final snapshot of model \texttt{RMHD\_P1\_6}, we also
computed the synthetic image at $\rm 0.3\,  \mu m$ in Fig.~\ref{fig:varscat}. At this wavelength, most of the surface brightness is due to star light scattered from the disk surface. The strongest
scattering happens at the rim, producing the ring
structure outward of $0.5$ AU. The spiral structure due to the vortex
is also visible. At larger radii ($R>1$ AU), the
intensity drops by orders of magnitude and a shadow is
visible. The vortex increases locally the inner rim height, throwing a non-axisymmetric shadow onto the outer
disk. Almost all the scattered 0.3-$\mu$m flux of the system comes from the inner rim. As recently discussed by \citet{don15}, the shadow structure
shows a smooth profile, without any sharp transitions. Although we do not expect such a
structure to be observed given the spatial resolutions that can be
reached with current telescope facilities, such shadowing by an inner
vortex might affect the disk temperature (and therefore its dynamical
response) through its impacts on heating and cooling. 

\begin{figure}
  \resizebox{\hsize}{!}{\includegraphics{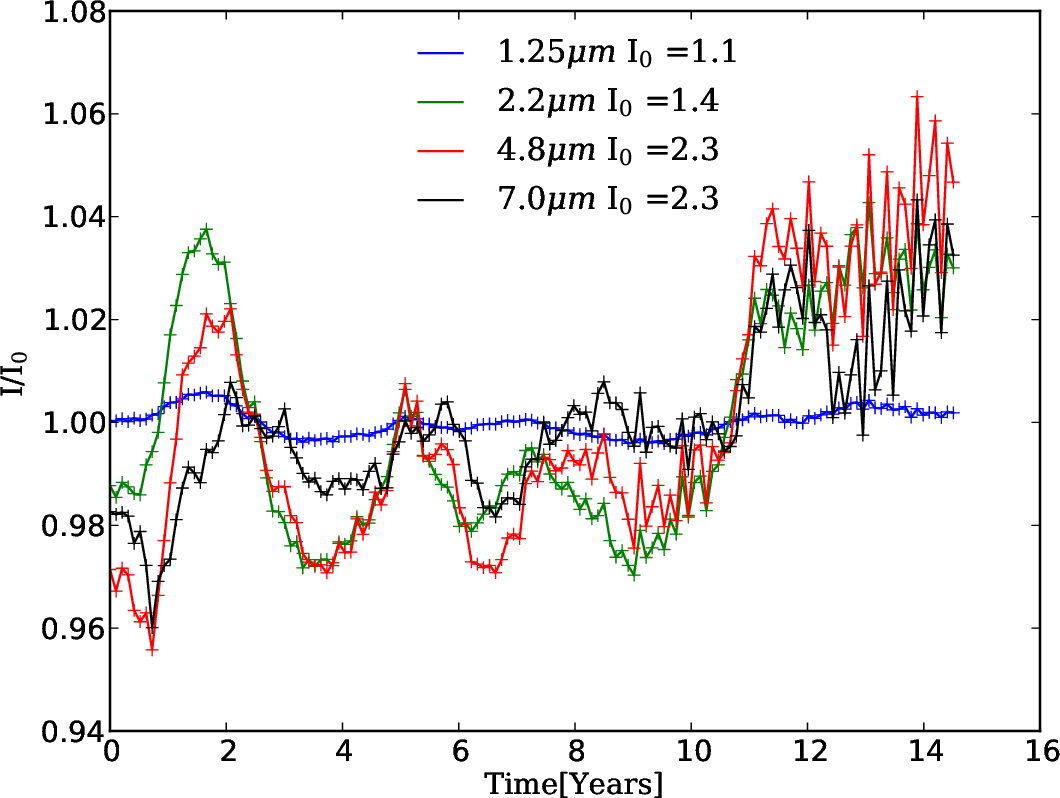}}
  \resizebox{\hsize}{!}{\includegraphics{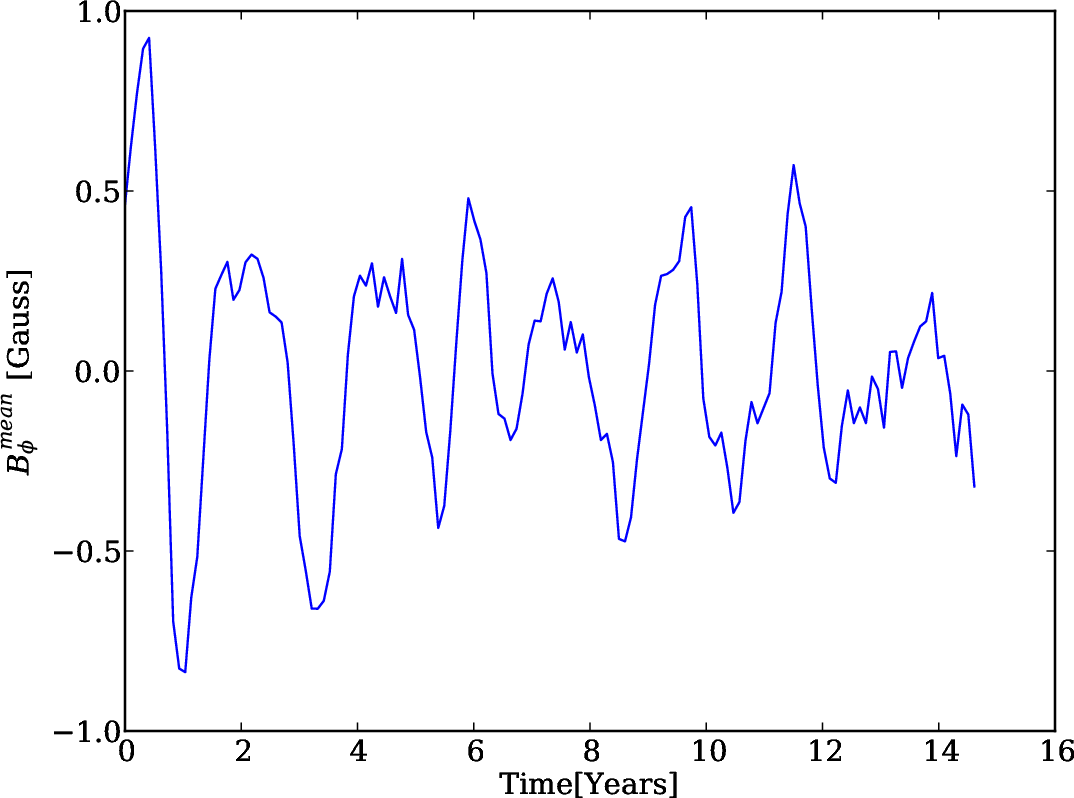}}
\caption{Top: Intensity over time for $\rm 1.25 \mu m$ (blue), $\rm 2.2 \mu m$ (green), $\rm 4.8 \mu m$ (red) and $\rm 7.0 \mu m$ (black) at 45.0$^\circ$ inclination from model \texttt{RMHD\_P1\_6}. The legend lists the normalizing intensities in units of $10^{-18}$ erg cm$^{-2}$ s$^{-1}$ Hz$^{-1}$. Bottom: Time evolution of the mean toroidal magnetic field, determined at a radius of 0.5 AU and a height of Z/R=0.05. Two years roughly corresponds to $10$ Keplerian orbits at 0.5 AU.} 
\label{fig:varem}
\end{figure}

\subsection{Variability}
\label{sec:var}

We now investigate potential variability using the results of model \texttt{RMHD\_P1\_6}. We focus on two aspects. The first is the variability of
the NIR emission caused by variations of the rim surface and
shape. The second is related to the star's occultation by the inner 
rim. 

\subsubsection{NIR intensity variability }

We start with the variability of the NIR emission, which we calculate
for a system viewed with an inclination of 45$^\circ$. The lightcurve is plotted in Fig.~\ref{fig:varem} for different wavelengths. At $\rm
1.25 \mu m$, the variations are smaller than
$1\%$. At this wavelength, most of the emission is coming from the
optically thin and hot material close to the midplane position of the inner rim. Larger
variations become visible at the wavelengths $\rm 2.2 \mu m$ and $\rm
4.8 \mu m$, reaching up to $\pm 5\%$ in relative amplitude. At longer
wavelengths, e.g. at $\rm 7 \mu m$, the fluctuations amplitude
decreases again. Fig.~\ref{fig:varem} also shows clearly that the
variations at different wavelengths are correlated.

The NIR emission exhibits two different types of
variations. During the early evolution ($0$-$10$ years),
Fig.~\ref{fig:varem} shows low frequency variations with a period of two years (which roughly corresponds to $10$ Keplerian orbits
at 0.5 AU). These variations could be connected to the oscillations of
the mean toroidal magnetic field which happen to display a similar timescale of $10$ orbits (see Fig.~\ref{fig:varem},
bottom panel, where we plot the mean toroidal magnetic field at a
radius of 0.5 AU and a height of Z/R=0.05). Oscillations of the mean
toroidal magnetic field such as reported here are known to be a robust
outcome of MRI-driven MHD turbulence in disks
\citep{sto96,mil00,les08,gre10,sim11,flo11b}. Our results thus potentially
suggest an indirect signature of this dynamical feature in the NIR
emission variability of Herbig stars. At later time during the
simulation ($t>10$ years), the frequency of the variability changes
and the NIR emission starts to display monthly time variations. These are due to the vortex rotating around the star and depend on the combination of
vortex azimuthal angle and disk inclination. 
%% The mean toroidal magnetic field shows also a strong initial oscillation,
%% followed by more regular oscillations on a roughly 2 year timescale
%% which matches the 10 orbital rotations at this location.  

\begin{figure}
  \resizebox{\hsize}{!}{\includegraphics{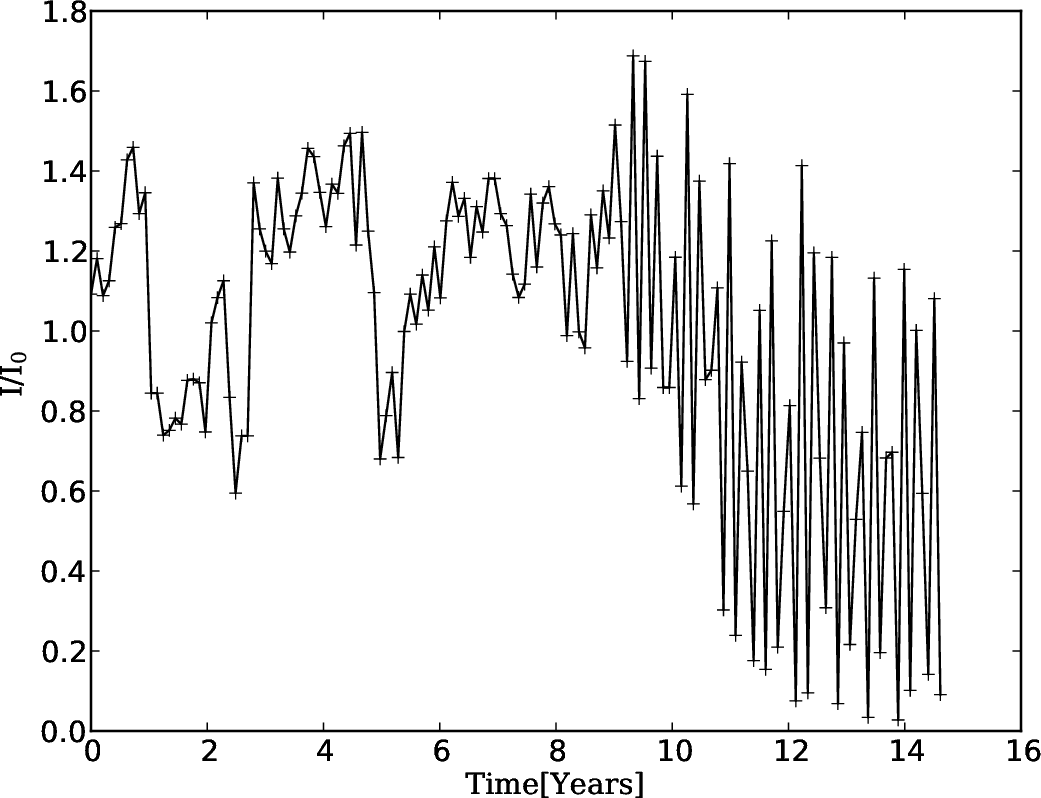}}
  \vspace{3mm}
  
  \resizebox{\hsize}{!}{\includegraphics{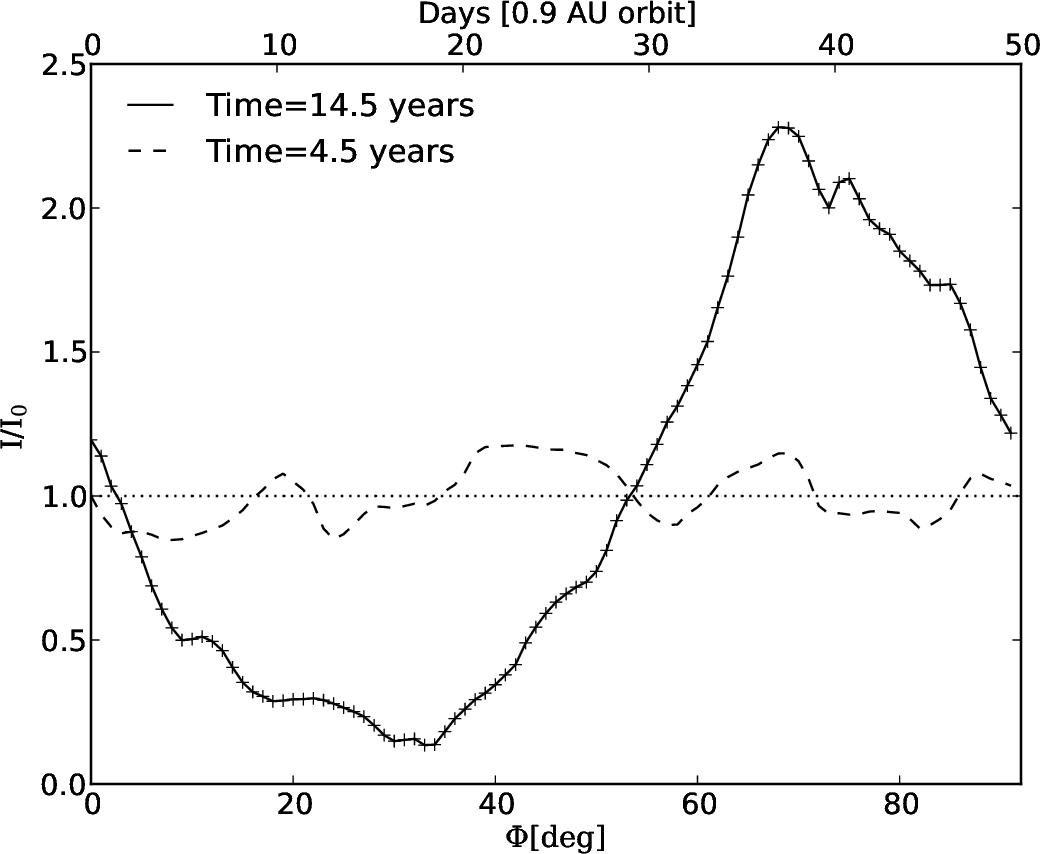}}
\caption{Top: Intensity over time at $\rm 0.3\,  \mu m$ and at 81.1$^\circ$ inclination of the global 3D radiation non-ideal MHD model \texttt{RMHD\_P1\_6}. The dips are caused due to variations in the dust density at the inner rim surface. 
Bottom: Finer sampling of the intensity over $\Phi$ angle at $\rm 0.3\,  \mu m$ and at 81.1$^\circ$ inclination using snapshots for two different times. The upper time axis corresponds to the rotation at 0.9 AU, the position of the occulting rim.} 
\label{fig:sed_var}
\end{figure}

\subsubsection{Stellar occultation by the inner rim}
Finally, we investigate the variability associated with the occultation of the star due to variations, both in space and time, of the inner rim height. Such height variations will produce a time-varying
absorption of the stellar light which we study here. To do so, we calculate a time series of the relative intensity (i.e., normalized by the mean intensity) received by an observer at $\rm 0.3\, \mu m$ when viewing the disk at an inclination $\theta=81.1^\circ$. We chose that particular inclination because this is the value of $\theta$ for which the variability is the largest. The wavelength at $\rm 0.3\, \mu m$ was chosen to represent the variations at the peak emission flux for this type of star.
 
The time evolution of the $\rm 0.3\, \mu m$ intensity is shown in Fig.~\ref{fig:sed_var} (top panel), with a time sampling of $0.1$ years ($\sim 1/6$ of an orbital period at 1 AU for this type of star). The amplitude of the variability is initially (i.e at times $t<10$ yrs) of the order of 50\% and displays an irregular pattern. It is caused by the turbulent motions at the rim surface. In addition, the mean field oscillations reported in Fig.~\ref{fig:varem} are also able to increase the density along the line-of-sight for a given time. At later times during the evolution ($t>10$ yrs), the vortex causes the variations to become more periodic with intensity fluctuations up to an order of magnitude.
To obtain a finer sampling of both occultation patterns, we make two additional series of Monte Carlo radiative transfer calculations, one for a representative snapshot from the pre-vortex stage of the disk's evolution, and the other for the later stage. For each, we simulate the time changes by systematically increasing the azimuthal angle from which we view the disk. The results are shown in
Fig.~\ref{fig:sed_var} (bottom panel). At $t=4.5$ yrs (dashed curve), we find variations of the order of $20 \%$ with a corresponding timescale of about ten days. They are due to turbulent structures
located at the rim surface and correspond to the high frequency fluctuations seen at early times on the top panel of Fig.~\ref{fig:sed_var}. At $t=14.5$ yrs (solid curve), we recover the periodic modulation of the
intensity, with an amplitude of roughly one order of magnitude, seen at late times on the top panel of Fig.~\ref{fig:sed_var}. As discussed previously (see the variations of the $\tau_*=1$ line of Fig.~\ref{fig:dustsl}), this is due to the rim height locally increasing by about $10\%$ at the location of the vortex.

\begin{figure}
\resizebox{\hsize}{!}{\includegraphics{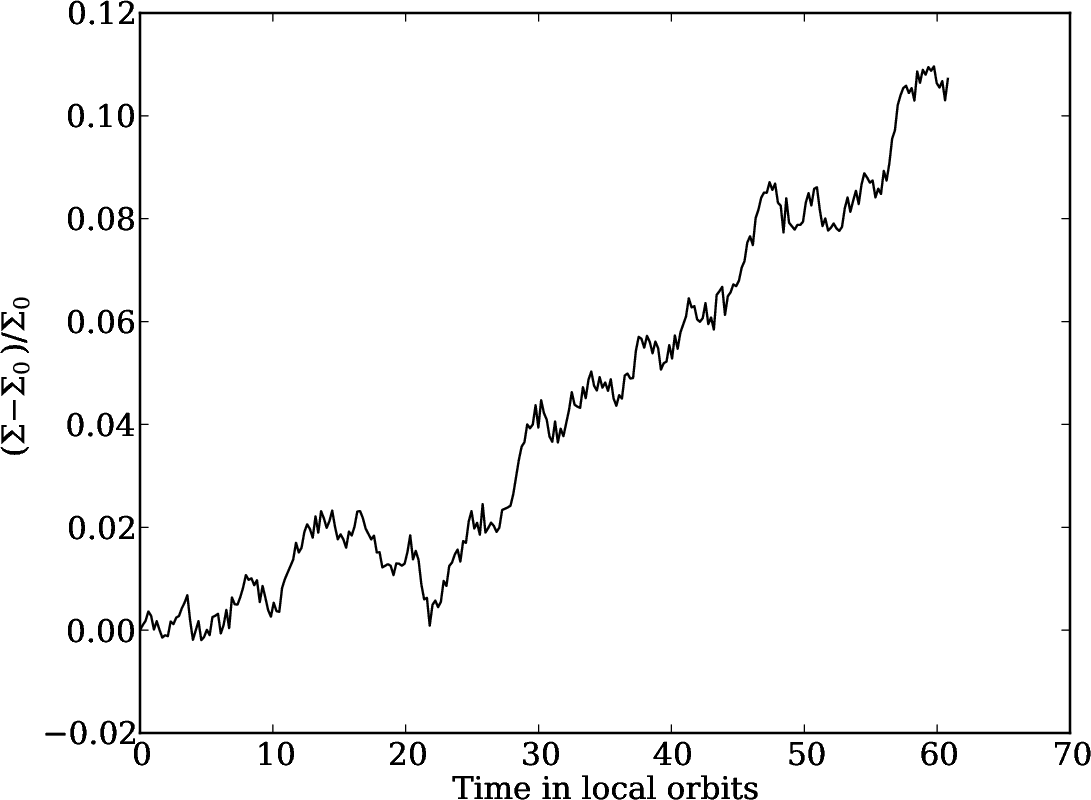}}
\caption{Evolution of the surface density over time shown at the dead-zone inner edge (0.85 AU) for model \texttt{RMHD\_P0\_4}.} 
\label{fig:sig_eq}
\end{figure}

\section{Discussion}
\label{sec:discussion}

In this section, we discuss some limitations of our modeling.

\subsection{Zero-net-flux models}

In section~\ref{sec:dyn} we have shown that the disk structure remains stable for the simulation runtime. However, the accretion stress inside the dead zone which is found for model \texttt{RMHD\_P0\_4} and model \texttt{RMHD\_P1\_6} is orders of magnitude lower than what was assumed to generate the initial conditions. The value of the surface density inside the dead-zone should therefore be taken with care. Fig.~\ref{fig:sig_eq} shows the surface density evolution over time at the dead-zone inner edge at 0.85 AU for model \texttt{RMHD\_P0\_4}. There we observe the fastest surface density variation. The surface density at the dead-zone inner edge increases around 10\% over the runtime. Over the accretion flow timescale we would expect a gradual increase of the surface density in the dead zone. Spanning such timescales is not feasible in 3D simulations.

\subsection{Vertical net flux model}
Model \texttt{RMHD\_P0\_4\_BZ} shows a much higher accretion stress in the dead zone, close to the
value assumed in computing the initial conditions. However, one should be
careful in interpreting the magnetic field strength and geometry at
large radii in this model given the short simulation time. 
The time average of $50$ inner orbits represents only
$\sim 1.6$ outer orbits and the high stress we observe in the
dead-zone might only be due to the initial conditions. Model \texttt{RMHD\_P0\_4\_BZ} is also limited by severe mass loss
associated with a strong magnetically driven outflow. Local and global simulations by \citet{fro13} and \citet{suz10} suggest that the
surface density would decrease on timescales of $\sim 100$ local
orbits as a result. This will eventually also affect the optical depth
and the thermal structure at the rim, setting a practical limit to the
maximum integration time of model \texttt{RMHD\_P0\_4\_BZ}.

\subsection{Ohmic resistivity}
We here treat only the Ohmic resistivity, though the other non-ideal 
effects -- ambipolar diffusion and Hall drift -- are important in 
the dead zone \citep{bai13,tur14b,les14}. We have compared Eq. \ref{eq:rem} which mimics the Ohmic dissipation dependence on temperature with the ionization balance results from \citet{des15}. Eq. \ref{eq:rem} reproduces the transition region around Elsasser number unity very well. However we note that the Ohmic resistivity is much greater inside the dead zone ($Re_m \ll 1 $) than our computational resources allow. Furthermore, we consider only thermal ionization, neglecting the stellar FUV and X-ray photons and the stellar and interstellar energetic protons. Thus the disk's surface layers, where these external fluxes ought to be absorbed, are ionized too little in the colder parts of our models. Treating these additional factors would likely affect the geometry of the magnetic fields in and around the dead zone.

\subsection{Comparison to previous works}
\citet{tur14a} showed that a magnetically supported atmosphere could
increase the near infrared excess emission by a factor of
two. However, the sublimation curve (Eq.~\ref{eq:ev}) was not fully treated in that paper. In our
models we find the rim's shape accurately using
Eq.~(\ref{eq:ev}). Most of the emission in the J and K band is then
coming from a narrow region which is not much affected by magnetic
fields, except in the outer parts of the rim. As a result, we find the NIR flux increase of about $20\%$ at $2$ microns when adding a strong and uniform vertical magnetic field. 
An even stronger magnetized corona would be needed to explain the NIR excess.

%\citet{hub09} reported
%magnetic fields for several Herbig stars while younger stars also show
%evidence for strong magnetic fields {\bf (refs?)}. {\bf SEB: These
%  fields are stellar 
%  fields, and likely create a magnetospheric cavity. It is not clear
%  that they correlate with the disk B-field. Are we saying
%  that the solution to this problem should lie in magnetospheric
%  cavities?} 
%Further work should include more realistic grain
%sizes and different compositions and grain species. We remind again
%that the grain size should be strongly reduced accross the inner
%rim. We observe high turbulent velocities, which would limit
%substantially the growth of dust particles \citep{bra08,bir12} in the
%region between the inner rim and the dead-zone inner edge ($\sim$
%between 0.5 and 0.8 AU). {\bf SEB: I don't understand how more
%  realistic grain size distribution would help: if anything, we would
%  have more larger grain size than we do now, and this would makes the
%problem even more difficult to solve, or are we including already a
%grain size distribution that we think would be even more biased toward
%small particles if we were to treat collisions properly?...We need to
%be clearer here? I am confused, and I think any reader would be...}

\subsection{Observed variability}
%
%Observations devoted to the issue of young star disk system
%variability has been the focus of intense research in the past decades
% %The recent surveys by the CoRot and Spitzer
%telescopes focused on the young stellar object variability (YSOVAR) of
%the T Tauri star systems 
%
%. They reported variations in the accretion rate by line
%emission and from the dust emission in the NIR on monthly timescales. 
%The reported timescales for the variations observed in Herbig systems matches very well 
%to our models.
%
%Those observations by \citet{sit12} found variations in the NIR bands in the
%order of 10 to 20\%. Our models predict variations by the MRI turbulence below 10\%. 
%
%\citet{wag15} investigated two states on the Herbig system HD 169142 on a timescale of 10 years apart. They reported a variability in the NIR of 45\% which they explained by a structural change of the outer edge of the inner dust rim. 
%
%Those observations indicate that there could be an addition mechanism causing those levels of variations, like a disk wind \citep{ban12} which could be driven by the MRI \citep{miy16}. 
%
%Further relations of the variability and the inner rim shape was found in scattered light observations of HD 163296 by \citet{wis08}, in which they report time variable self-shadowing at the inner rim on timescales of several years. 

Variability of young star-disk systems has been the focus of intense research for decades
\citep{joy45,car01,mor11}. Large sets of month-long simultaneous optical and infrared lightcurves became available recently through surveys of low-mass T~Tauri stars with the CoRoT and Spitzer space telescopes \citep{cod14,sta16}. 
A few Herbig stars have been examined through imaging and/or spectroscopic time series.
 \citet{sit12} studied the variability of the gas and dust emission from the Herbig system
SAO 206462, reporting spectral line changes connected with variations in the accretion rate, and near-infrared dust flux changes on monthly timescales with amplitudes of 10 to 20\%. Our models match the observed timescales very well, but the MRI turbulence yields somewhat lower amplitudes below 10\%. \citet{wag15} investigated the Herbig system HD~169142 at two epochs separated by 10~years. They reported near-infrared variability of 45\%, explaining it by a structural change in the inner dust rim.
These observations suggest there may be an additional variability mechanism, perhaps associated with a disk wind \citep{ban12} which could be driven by the MRI \citep{miy16}. Further linking the variability to the inner rim shape are scattered light observations of HD~163296 by \citet{wis08}. They report that shadows cast by the inner rim vary on timescales of several years. We finally note that occultation by vortices might be observed in highly inclined systems. An example is AA~Tau, which is inclined about $75^\circ$ and still undergoing a strong occultation event that began in 2013 \citep{bou13}. Such an event could be due to a local thickening of the disk at a vortex like that on the inner rim in the models we present, but located further from the star. Highly inclined disks appear well suited to observe the rim occulting the star, especially in cases where the outer disk is dust-depleted by radial drift or settling \citep{ber00}. Other studies relating flux dips in highly inclined systems to occultation by dusty material include those by \citet{ale10,mor11,cod14}.

%We finally note that the occultation by vortices could be observed in high inclined disk systems. One example is the system AA Tau, showing a still ongoing, strong occultation event in 2013 \citep{bou13}. This system has a high inclination of around $75^o$ \citep{bou13}. 

Such an event could be due to a vortex located at larger disk radii, increasing locally the height of the disk, similar as the vortex at the inner rim in the models we present. 

Such highly inclined disk systems might be ideally suited to observe the star occultation by the rim \citep{ber00}, especially in cases where the outer disk is dust depleted by the radial drift or settling. The studies by \citet{ale10,mor11,cod14} related dipping events by dust occultation which was observed in highly inclined systems. Recently \citet{ans16} showed that such events could also occur for less inclined systems.

\section{Summary}
We have presented the first global 3-D radiation non-ideal MHD models of the innermost reaches of protostellar disks, using them to investigate the dynamics and thermodynamics of the planet-forming material.  Our models include the transfer of the starlight into the dust and gas, where the heating impacts the dust sublimation and deposition and the Ohmic resistivity. The starting conditions come from axisymmetric radiation viscous hydrodynamical models of the disk around a typical Herbig~Ae star, with a radially-independent mass accretion rate of
$10^{-8}$~M$_\odot$~yr$^{-1}$.  Magnetic fields either with or without
a net vertical flux yield magneto-rotational turbulence.  The inner disk's structure divides naturally into four zones:

\begin{enumerate}

\item[1.] \hspace{0.5mm} Between the star and the silicate front, the gas is turbulent with RMS speeds of 400 to 800~m~s$^{-1}$, depending on the initial magnetic field configuration. The accretion stress-to-pressure ratio $\alpha$ is between 3 and 10\%, and the turbulent magnetic field strengths are several Gauss. The gas is hotter than the silicate sublimation threshold.

\item[2.] \hspace{0.5mm} Lower temperatures let silicate dust exist beyond a curved front that is closest to the star in the midplane at about 0.5~AU. Dust and strong turbulence coexist at 0.5 to 0.8~AU, where temperatures are about 1000~K.  The stress-to-pressure ratios are similar to those nearer the star, but the turbulence is slightly slower, due to the lower temperatures, at 300 to 700~m~s$^{-1}$.  High-speed collisions should substantially limit the grains' maximum size.

\item[3.] \hspace{0.5mm} Beyond about 0.8~AU, temperatures are low enough and collisional ionization slow enough that the magnetic fields decouple from the gas motions. This region is the dead zone.  From 0.8 to 1.1~AU, turbulent speeds decline quickly with distance. Density waves propagating from the turbulent region are quickly damped, leaving laminar gas with turbulent velocities below 1~m~s$^{-1}$. A local pressure maximum lies in the weakly-turbulent dead zone near 1~AU. This pressure peak is able to halt solid particles' radial drift.

\item[4.] \hspace{0.5mm} Beyond 1.1~AU, well inside the dead-zone, the disk is quasi-laminar with very low turbulent speeds.  This region lies partly in the shadow cast by the sublimation front.

\end{enumerate}

The 3-D calculations let us investigate non-axisymmetric stability.
We find Rossby wave instability develops over timescales of 20 local
orbits into a vortex located at the dead zone's inner edge and close
to the upper rim of the curved silicate sublimation front.  The vortex
moves the disk's surface up and down over its orbital period.

We post-process our results using Monte Carlo radiative transfer tools
to compare against a variety of observational constraints:

\begin{enumerate}

\item \hspace{0.5mm} Our models with strong magnetic fields have near-infrared fluxes that are 5 to 20\% greater than the viscous version, because the magnetically-supported disk atmosphere raises the sublimation front, reprocessing more of the starlight into wavelengths near 2 and 3~$\mu$m. Magneto-rotational turbulence could thus be a factor in the long-standing puzzle of Herbig stars' anomalously large near-infrared excesses \citep{vin06,ack09,dul10}.

\item \hspace{0.5mm} The vortex that develops near the sublimation front's high point locally raises the height where the starlight is absorbed, thus casting a longer shadow on the disk beyond. The fraction of the stellar luminosity intercepted by the sublimation front at this stellar longitude is increased about 10\%.  Such shadow-casting vortices could potentially be related to the variability observed in scattered-light imaging of Herbig disks \citep{wis08}.

\item \hspace{0.5mm} The near-infrared flux varies up to 10\% due to the movements of the inner rim, on timescales of months to years. The regular component of the variations is larger relative to the irregular component when the vortex is present.  Further development of this picture could help in understanding why young stars with protostellar disks have such diverse optical and infrared lightcurves \citep{sit12,cod14}.
\end{enumerate}

Radiation-MHD models of the kind we have demonstrated here open a new window for investigating protoplanetary disks' central regions.  They are ideally suited for exploring young planets' formation environment, interactions with the disk, and orbital migration, in order to understand the origins of the close-in exoplanets.

\section*{Acknowledgments}

We thank John Stauffer for his valuable comments on the manuscript. We thank also Satoshi Okuzumi for helpful discussions during the project. We thank Andrea Mignone for supporting and advising us with the newest PLUTO code. Parallel computations have been performed on the Genci supercomputer 'curie' at the calculation center of CEA TGCC and on the zodiac supercomputer at JPL. For this work, Sebastien Fromang and Mario Flock received funding from the European Research Council under the European Union's Seventh Framework Programme (FP7/2007-2013) / ERC Grant agreement nr. 258729. 
This research was carried out in part at the Jet Propulsion Laboratory, California Institute of Technology, under a contract with the National Aeronautics and Space Administration and with the support of the NASA Exoplanet Research program via grant 14\-XRP14\_2\-0153. Copyright 2016 California Institute of Technology. Government sponsorship acknowledged.

\appendix

\section{{\bf A} Comparison $\Delta \phi=0.4$ vs. $\Delta \phi=1.6$}
\label{app:comp}
\begin{figure}%[!htb]
%\begin{minipage}{0.4\textwidth}%
%\includegraphics[width=\linewidth]{FIG/ALPHA1D_comp.png}%
%\endminipage\hfill
%\end{minipage}\hfill%
%\begin{minipage}{0.4\textwidth}%
%\includegraphics[width=\linewidth]{FIG/densvar_m.png}%
%\endminipage\hfill
%\end{minipage}\hfill%
%
  \resizebox{0.5\hsize}{!}{\includegraphics{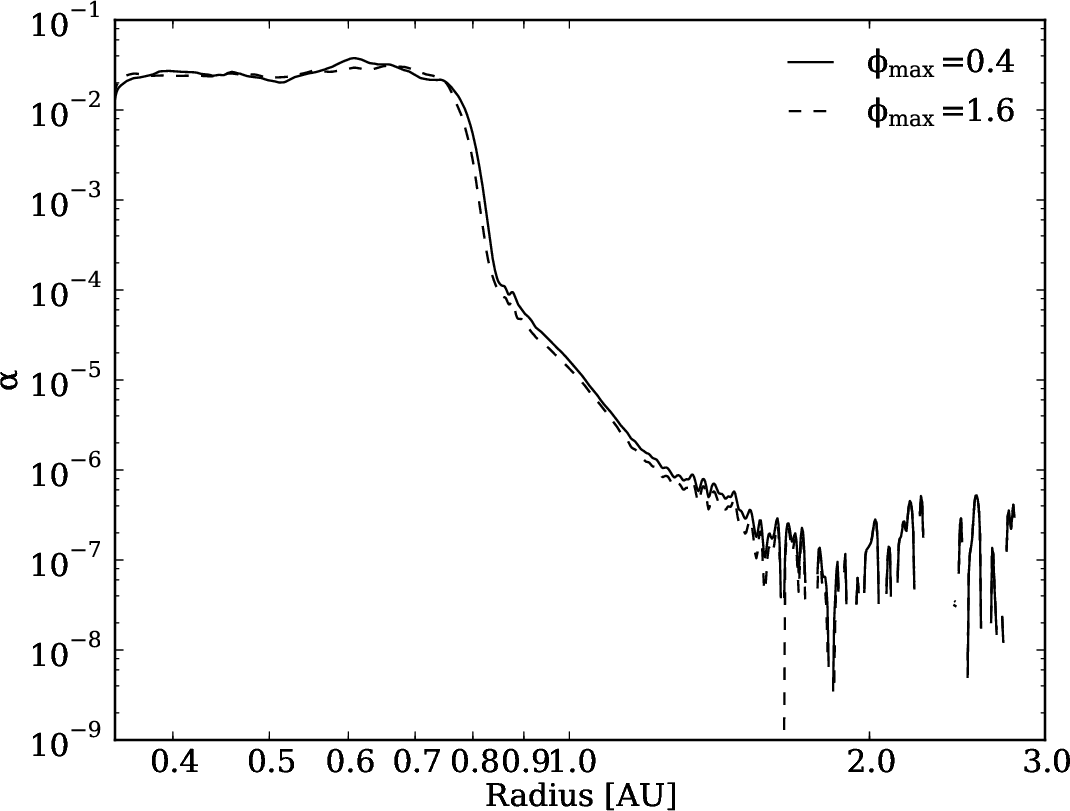}}
  \resizebox{0.5\hsize}{!}{\includegraphics{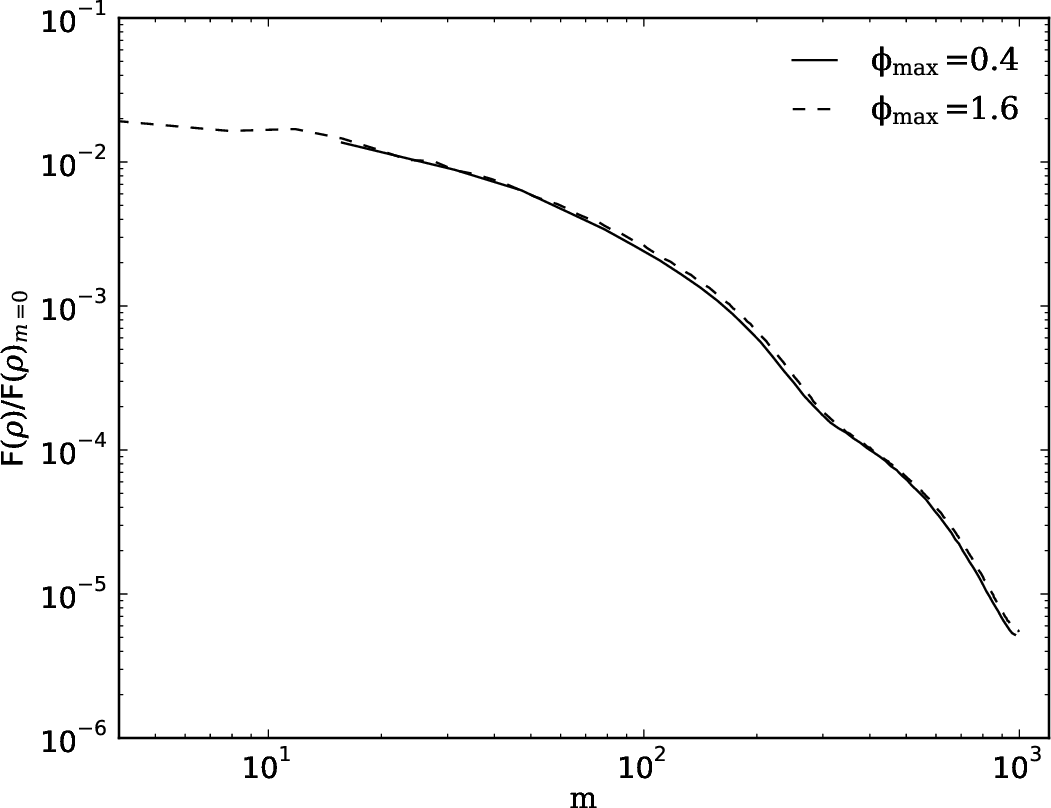}}
\caption{Left: Time averaged radial profile of the stress to pressure ratio $\alpha$ for model \texttt{RMHD\_P0\_4} (solid line) and model \texttt{RMHD\_P1\_6} (dashed line). Right: Fourier transfer of the midplane density in azimuth at the midplane at 0.5 AU for model \texttt{RMHD\_P0\_4} (red dashed line) and model \texttt{RMHD\_P1\_6} (solid line). \label{fig:al_comp}}%
\vspace{2mm}
\end{figure}

We previously reported stronger accretion stress in a smaller azimuthal domain due to stronger mean fields \citet{flo12}. Here we compare the two models with different azimuthal domains. In Fig.~\ref{fig:al_comp}, left, we show the radial profiles of $\alpha$ in the models \texttt{RMHD\_P0\_4} and \texttt{RMHD\_P1\_6} averaged over the same time period of 50 inner orbits. The two models present similar radial profiles with no significant differences. However, we note that we consider here a zero-net-flux field while a net-flux field was used in our previous simulations \citep{flo12}. In addition, we think that the short azimuthal domain still can be an issue for the model \texttt{RMHD\_P0\_4\_BZ}, which uses a net vertical flux field. Here, the mean field becomes nearly as large as the turbulent component (Section~\ref{sec:vert}) which could be a sign that the azimuthal domain is too small.

In Fig.~\ref{fig:al_comp}, right, we take a final look at the density perturbations of model \texttt{RMHD\_P0\_4} and \texttt{RMHD\_P1\_6}. Here, we calculate the Fourier transform of the density along azimuth at the midplane at 0.5 AU using the same time average as before. Fig.~\ref{fig:al_comp}, right, shows that the profiles look very similar, from which we conclude that both models represent similar turbulent characteristics.

\section{{\bf B} Generating initial magnetic field configurations}
\label{ap_1}
In the following steps we explain how to generate the initial random magnetic field. 
Such a field has the advantage that the MRI turbulence quickly reaches a steady-state.
\begin{enumerate}
\item $\, $Calculate a random vector potential for the component $\rm A_r=A^{rand} f_{r} f_{\theta} r$ with r being the radius, $\rm f_{r}$ being a parabolic damping factor proportional to $(r -r_0)^2$ to set $\rm A_r=0$ at the inner and outer radial boundary. $\rm f_{\theta}$ is a parabolic factor which decreases the potential from the midplane to the $\rm \theta$ boundary by the factor 50 to account for the decrease of the magnetic field in the disk corona. 
\item $\, $The amplitude of the vector potential $\rm A_r$ is set to match a plasma beta value of 100 at the midplane.
\item $\, $Apply a Gaussian filter ($\rm \sigma=2 \delta r$) to smooth out fluctuations on grid level to obtain comparable scales as the turbulent field in the steady-state.
\item $\, $Calculate $\rm \nabla \times A$ to obtain an initial $\rm B_{\theta}$ and $\rm B_{\phi}$.
\end{enumerate}
It is important to define the vector potential with a periodic boundary condition in $\rm \phi$ direction. Otherwise there will be a violation of the $\nabla \cdot B=0$ condition at the $\phi=0$ boundary. Initially the amplitudes of $\rm B_{\theta}$ and $\rm B_{\phi}$ are equal. A snapshot of the initial magnetic field is shown in Fig. \ref{fig:hdinit}.

For model \texttt{RMHD\_P0\_4\_BZ}, we add to the vector potential a constant value of $\rm A_\phi$. The value of $\rm A_\phi$ is chosen to match a magnetic field strength of 100 mGauss at 1 AU, which corresponds to a plasma beta $\rm \beta=2 P/B_z^2$ of $\rm \beta=3.5 \cdot 10^{4}$ at 1 AU at the midplane. The resulting vertical field has a radial profile of $\rm r^{-1}$. The strength of the field corresponds to a relative high value of vertical magnetic flux \citep{oku14}.

\section{{\bf C} Wave damping at the dead-zone inner edge}
\label{sec:damp}
There are several damping mechanisms for the density waves, generated in the MRI turbulent regions and which are traveling into the dead-zone. The most important one is the non-linear damping by shocks as soon as the wavelength is comparable to the disk scale height \citep{hei09b}. A similar wave dissipation by weak shocking in the dead-zone was found in global MHD simulations at the inner dead-zone edge by \citet{fau14}. One difference, to previous simulations is the fact that the local H/R is much smaller, meaning that waves traveling over a given radial distance are stronger damped. Another difference is the radial changing thermal diffusion.  The density waves travel through a region with increasing surface density, and so increased optical thickness. At the same time, in this region the MRI is switched of and there is no excitation of density waves anymore. 

\begin{figure}
  \resizebox{\hsize}{!}{\includegraphics{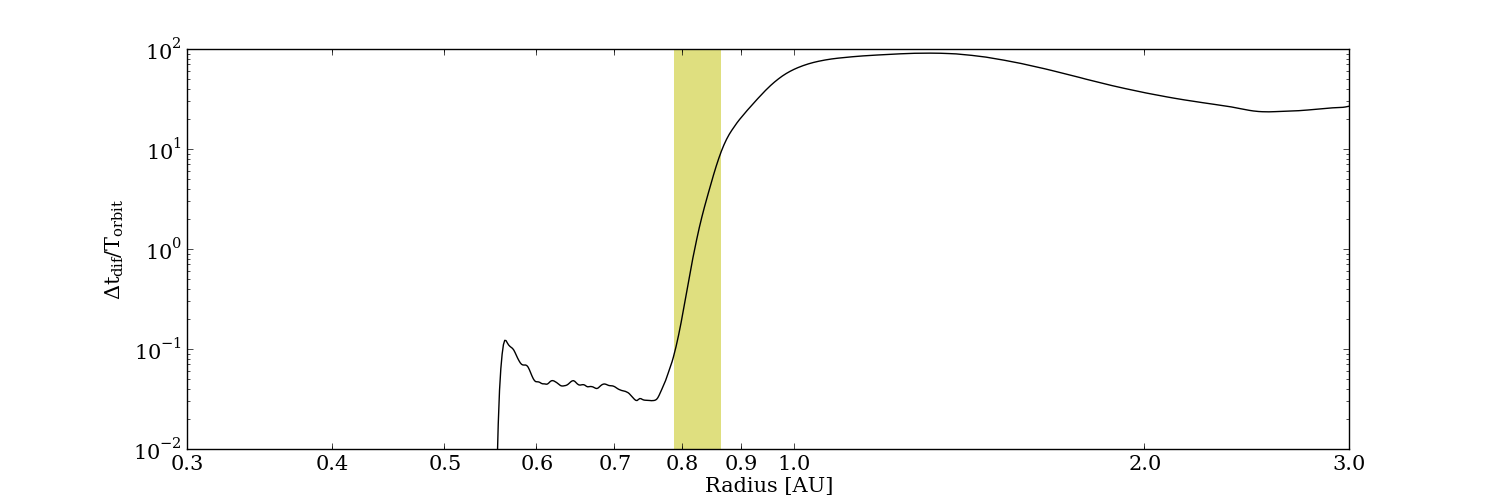}}
\caption{Thermal diffusion timescale along the midplane for the snapshot at 50 inner orbits for model \texttt{RMHD\_P1\_6}. The profile follows roughly the profile of the dust density which determines the opacity and so the local optical depth. The yellow bar marks the region in which the diffusion timescale becomes comparable to the dynamical timescale (0.1 to 10 times the orbital time), for which we expect highest damping.}
\label{fig:dt_diff}
\end{figure}

The efficiency of wave damping by thermal diffusion is highest if the diffusion timescale is comparable with the typical timescale of the density wave. The turbulent correlation time of the MRI is roughly one tenths of the orbital period for lengthscales comparable to H. At the same time, \citet{hei09b} report that the largest density waves fitting in the azimuthal domain carry most of the energy. The characteristic timescale of thermal diffusion can be estimated with the radiation diffusion in the gas pressure dominated regime \citep{fla10}:

\begin{equation}
\rm \Delta t_{dif}= \left (1 + \frac{\rho \epsilon}{4 E_R} \right ) \frac{3 \kappa \rho L^2}{c}
\end{equation}
with the characteristic lengthscale L, which we set to the disk scale height H.
In Fig.~\ref{fig:dt_diff} we plot the midplane diffusion timescale for this lengthscale, normalized over the dynamical timescale. The yellow bar marks the region in which the diffusion timescale becomes comparable to the dynamical timescale, for which we expect highest damping. The zone matches the region in which the Reynolds stress quickly drops.
 
We summarize that the combination of a small H/R and a thermal diffusion timescale comparable to the dynamical timescale at the dead-zone inner edge leads to an efficient damping of the density waves and so the Reynolds stress in the dead-zone. 

\section{{\bf D} RT setup and dust halo} 
\label{sec:ap3}

We briefly summarize the setup for RADMC3D to post-process the 3D datasets. For the Monte-Carlo runs we use 210 million photon packages. We first transfer the grid values from the radiation MHD calculation to the RADMC grid structure. Then we re-calculate the thermal structure using the wavelength dependent dust opacity table. This ensures that the SED and the temperatures are consistent for the given dust opacity. For the dust opacity we assume the same grains size distribution as for the 2D models (Appendix A \citep{flo16}). All grains have the same size distribution, including those in the dust halo in front of the dust rim. In reality, larger grains would be more likely to survive the hotter temperatures in front of the inner rim due to their higher emission to absorption ratio \citep{kam09}.

We have checked that the temperature of the Monte Carlo run and the global RMHD models match exactly at the inner rim. As we neglect the gas opacity in the Monte-Carlo runs, we observe small deviations in the temperature for the very optical thin layers of the global models. In addition, we have already shown in our previous models that the effect of the accretion heating remains small for these model parameters \citep{flo16}.

We want also to discuss again the dust halo which appears in our models \citep{flo16}. The main reason for this halo is the difference between the gas and dust temperatures in the optically thin environment. For the case $T> T_{ev}$, the dust starts to evaporate, however pure gas alone would lead to a temperature below the evaporation temperature. The solution is that a tiny amount of dust condenses to balance the temperature drop. The result is a small dust halo in which the temperature is close to the evaporation temperature. A similar result was found by \citet{kam09}, however in their model, larger dust grains are responsible as they have a larger emission to absorption ratio and so survive in front of the rim. In our models, the gas component has a larger $\epsilon$ value which leads to a temperature which is cooler than dust in an optical thin environment.

\section{{\bf E} MRI quality factor} 
\label{sec:ap4}

\begin{figure}
\centering
  \resizebox{0.5\hsize}{!}{\includegraphics{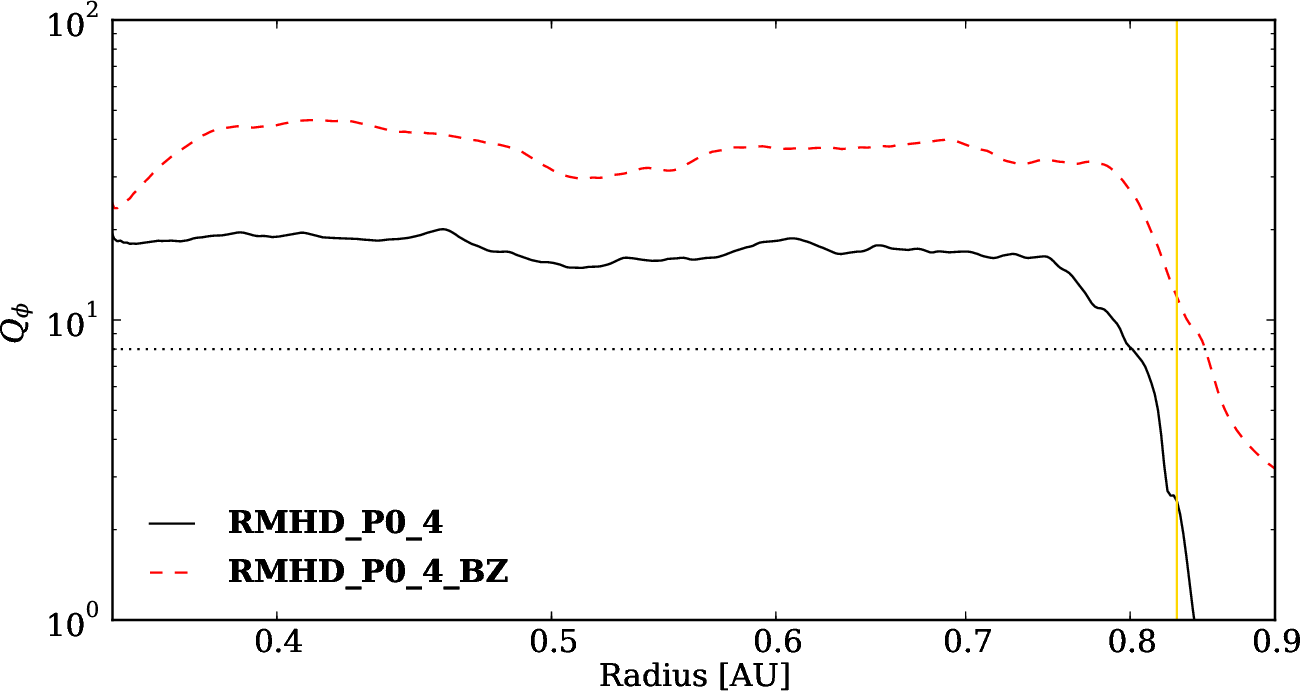}}
\caption{The quality factor, time averaged and mass weighted in the vertical direction. The dead-zone edge is annotated with the yellow line. The black dotted lines emphasize the 8 cell limit.}
\label{fig:quafa}
\end{figure}

Following the work by \citet{nob10} and \citet{sor11} we determine the quality factor Q which shows the number of grid cells per fastest MRI growing mode. The quality factor $Q_\phi$ for the azimuthal field is defined as 
\begin{equation}
Q_\phi = 2 \pi \sqrt{\frac{16}{15}} \frac{|B_\phi|}{\sqrt{4 \pi \rho}} \frac{1}{\Omega } \frac{1}{r \Delta \phi}.
\end{equation}
In Fig.~\ref{fig:quafa} we plot the quality factor for the models \texttt{RMHD\_P0\_4} and \texttt{RMHD\_P0\_4\_BZ}. Space and time averaging follows the same strategy as in Section~\ref{sec:timeav} and ~\ref{sec:vert}. The plot shows that both models resolve very well the MRI in the region with high ionization with 16 or more grid cells per fastest-growing MRI wavelength. Model \texttt{RMHD\_P0\_4\_BZ} shows a even higher quality factor due to the stronger field (Figs.~\ref{fig:turbq} and~\ref{fig:turbq_vert}).

%\section{{\bf E} Calculating the height of the inner rim} 
%\label{sec:ape}

%To study the shadowing efficiency by the inner rim we first determine accurately the height of the rim. This is done by determining the height at which the star is getting occulted. We determine the height of occultation for the stellar irradiation at $\rm 0.3\,  \mu m$ at an inclination of $81.1^\circ$ for model \texttt{RMHD\_P1\_6} after 50 inner orbits. At this inclination, the intensity drops by a factor of $e$. To increase the accuracy, we excluded the outer disk beyond 1 AU to avoid effects from the outer disk. By doing this we are able to measure the star occultation efficiency by the inner rim surface alone. Fig.~\ref{fig:phi_inc}, shows the intensity for different inclinations at two different $\Phi$ angles at minimum ($33^o$) and maximum ($68^o$) occulation for the latest snapshot of model \texttt{RMHD\_P1\_6}, including the full domain Fig.~\ref{fig:phi_inc}, left, and exluding the disk beyond 1 AU, Fig.~\ref{fig:phi_inc}, right.

%\begin{figure}
%  \resizebox{0.5\hsize}{!}{\includegraphics{FIG/VAR_PHI_INC_FULL.png}}
%  \resizebox{0.5\hsize}{!}{\includegraphics{FIG/VAR_PHI_INC.png}}
%  \caption{Inclination sample at minimum ($\Phi=33^o$) and maximum ($\Phi=68^o$) occulation for the latest output from model \texttt{RMHD\_P1\_6} for the full domain (left) and by excluding the domain beyond 1 AU (right).} 
%\label{fig:phi_inc}
%\end{figure}

\newpage
\bibliographystyle{aa}
\bibliography{IEDGE}

\end{document}